\documentclass[sigplan,screen]{acmart}

\startPage{1}
\setcopyright{rightsretained}
\acmPrice{}
\acmDOI{10.1145/3519939.3523437}
\acmYear{2022}
\copyrightyear{2022}
\acmSubmissionID{pldi22main-p84-p}
\acmISBN{978-1-4503-9265-5/22/06}
\acmConference[PLDI '22]{Proceedings of the 43rd ACM SIGPLAN International Conference on Programming Language Design and Implementation}{June 13--17, 2022}{San Diego, CA, USA}
\acmBooktitle{Proceedings of the 43rd ACM SIGPLAN International Conference on Programming Language Design and Implementation (PLDI '22), June 13--17, 2022, San Diego, CA, USA}

\usepackage[utf8]{inputenc}
\usepackage{hyperref}
\usepackage{microtype}
\usepackage{tabularx}
\usepackage{ragged2e}
\usepackage{amsmath}
\usepackage{subcaption}
\usepackage{mathpartir}
\usepackage{mathtools}
\usepackage{tensor}
\usepackage{sidecap}
\usepackage{enumitem}
\sidecaptionvpos{figure}{m}

\usepackage[english]{babel}
\addto\extrasenglish{
    
}
\addto\extrasenglish{
    
}

\usepackage{listings}
\usepackage{amsfonts}
\definecolor{keywordcolor}{rgb}{0.5,0,0.5}
\definecolor{textgray}{gray}{0.4}
\definecolor{mygray}{rgb}{0.5,0.5,0.5}
\lstdefinelanguage{none}{
    identifierstyle=
}

\DeclareCaptionStyle{mathpls}{
    labelfont={bf, small},
    labelsep=period,
    justification=RaggedRight,
    singlelinecheck=true,
}%

\title{DISTAL: The Distributed Tensor Algebra Compiler}



\author{Rohan Yadav}
\affiliation{%
    \institution{Stanford University}
    \streetaddress{353 Jane Stanford Way}
    \city{Stanford}
    \state{CA}
    \postcode{94305}
    \country{USA}
}
\email{rohany@cs.stanford.edu}

\author{Alex Aiken}
\affiliation{%
    \institution{Stanford University}
    \streetaddress{353 Jane Stanford Way}
    \city{Stanford}
    \state{CA}
    \postcode{94305}
    \country{USA}
}
\email{aiken@cs.stanford.edu}

\author{Fredrik Kjolstad}
\affiliation{%
    \institution{Stanford University}
    \streetaddress{353 Jane Stanford Way}
    \city{Stanford}
    \state{CA}
    \postcode{94305}
    \country{USA}
}
\email{kjolstad@stanford.edu}

\newcommand{\name}{DISTAL}
\definecolor{todocolor}{rgb}{0.8,0,0}
\newcommand{\TODO}[1]{{\color{todocolor}#1}}
\newcommand{\ignore}[1]{}
\newcommand{\mT}{\mathcal{T}}
\newcommand{\mM}{\mathcal{M}}
\newcommand{\mP}{\mathcal{P}}
\newcommand{\mF}{\mathcal{F}}

\newcommand{\mD}{\mathcal{D}}
\newcommand{\mI}{\mathcal{I}}
\newcommand{\pluseq}{\mathrel{+}=}
\newcommand{\mapscript}[3]{%
    \operatorname{%
            {\vphantom{#2}}_{#1}%
        \kern-\scriptspace
        \mathnormal{#2}_{#3}%
    }%
}
\newcommand{\tdistname}[3]{#1\mapscript{#2}{\mapsto}{#3}\mM}
\newcommand{\tdist}[2]{\tdistname{\mT}{#1}{#2}}

\newcommand{\bnfdef}{\mathrel{::=}}
\newcommand{\bnfalt}{\mathrel{\mid}}

\begin{document}

\begin{abstract}
    We introduce \name{}, a compiler for dense tensor algebra that targets modern
    distributed and heterogeneous systems.
    \name{} lets users independently describe how tensors and computation map onto
    target machines through separate format and scheduling languages.
    The combination of choices for data and computation distribution creates a large design space
    that includes
    many algorithms from both the past (e.g., Cannon's algorithm) and the present (e.g., COSMA).
    \name{} compiles a tensor algebra domain specific language to
    a distributed task-based runtime system and supports nodes with multi-core CPUs and multiple GPUs.
    Code generated by \name{} is competitive with optimized codes for matrix multiply
    on 256 nodes of the Lassen supercomputer and outperforms existing systems by between 1.8x to 3.7x
    (with a 45.7x outlier) on higher order tensor operations.
\end{abstract}

\maketitle

\section{Introduction}
\label{sec:intro}
Tensor algebra kernels are key components of many workloads that benefit from the compute,
memory bandwidth and memory capacity offered in a distributed system.
However, the implementation of distributed tensor algorithms that are both correct and achieve high performance is a
challenging task for most programmers.
The situation is even more daunting when taking into account the heterogeneity within a single
compute node; in many high performance systems today, managing the non-uniform memory access costs
between the multiple GPUs and CPU sockets within a node is another distributed systems
challenge to solve.

\begin{figure}
    \centering
    \includegraphics[width=\linewidth]{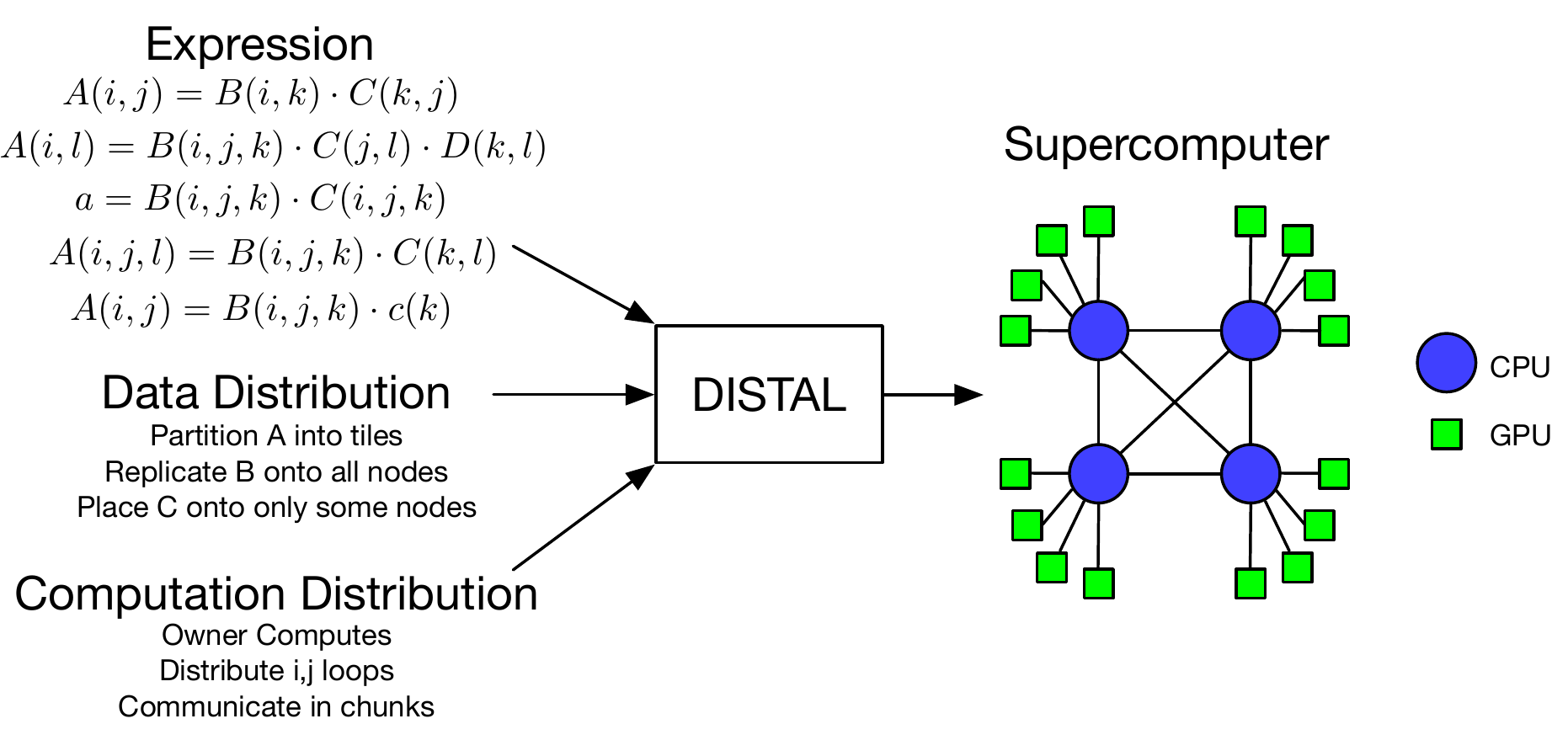}
    \caption{Example of the user-specified input (tensor algebra expressions, desired data and
    computation distribution) and output (code targeting supercomputers) of \name{}.}
    \label{fig:visual-example}
\end{figure}

We present \name{}, a compilation-based system that provides novel abstractions to create
implementations of any dense tensor algebra expression for modern heterogeneous machines.
\autoref{fig:visual-example} depicts how \name{} lets users
describe how the data and computation of a tensor algebra expression map onto a target machine.
\autoref{fig:lang-sample} shows C++ code that implements a multi-GPU distributed matrix-multiply using
\name{}.
Lines~\ref{fig:lang-sample:line:machine}--\ref{fig:lang-sample:line:dist-format} map tensors onto the machine
as part of the tensors' \emph{format}, and lines~\ref{fig:lang-sample:line:sched-begin}--\ref{fig:lang-sample:line:sched-end}
map computation onto the machine through a loop transformation-based \emph{scheduling language}.
By separating the specifications of data distribution and computation distribution, \name{} allows for their independent optimization, or for adapting either the data or computation distributions to complement the other.

Defining the distribution of both data and computation creates a design space of algorithms for
each tensor computation.
In particular, many algorithms from the literature are expressible
as data distributions and schedules in \name{}, including all of the algorithms
in \autoref{fig:algorithm-schedules} (\autoref{sec:algorithms}).
The abstractions of \name{} let these algorithms be expressed with the expected
asymptotic behavior and excellent practical performance.
In fact, our evaluation (\autoref{sec:evaluation}) demonstrates that implementations of these algorithms
using \name{} are competitive with hand-tuned implementations.

Implementations of tensor algebra operations on modern hardware can be hundreds
to thousands of lines of low-level code that manage data movement between nodes and
accelerators.
For example, a conservative estimate of the core distributed matrix-multiplication logic in the implementation of the COSMA~\cite{cosma} algorithm by the original authors is around 500 lines of code, excluding lower-level communication
code, local GEMM operations, whitespace, and comments.
In contrast, the full data placement and distribution related scheduling for a GEMM using
\name{} is 15 lines of code (in \autoref{fig:lang-sample}), while delivering competitive performance.

Alternatively to hand-coded implementations, state-of-the-art distributed tensor algebra libraries such as the Cyclops Tensor Framework (CTF)~\cite{ctf} achieve generality by
decomposing tensor algebra expressions into a series of distributed matrix multiplication and transposition
operations, relying on the efficiency of a hand-written set of core implementations.
These approaches cannot implement the best algorithm for every tensor expression,
as writing a hand-tuned implementation for every situation is impractical.
With \name{}, users can generate bespoke implementations of their tensor expressions that
implement either algorithms from the literature or new algorithms tuned to a target machine.

Additionally, tensor algebra kernels that run on a distributed machine do not exist in a vacuum.
These kernels operate on and generate data in the context of a larger application that imposes constraints on
how the program's data is partitioned and distributed among different memories in the target machine.
Libraries such as ScaLAPACK~\cite{scalapack} offer a set of kernels that assume a specific set of input distributions
and require the user to reorganize their data into one of these distributions, which can result in additional data movement.
In contrast, \name{} lets users specialize computation to the way that data is already
laid out, or easily transform data between distributed layouts to match the computation.

\begin{figure}
    \includegraphics[width=\linewidth]{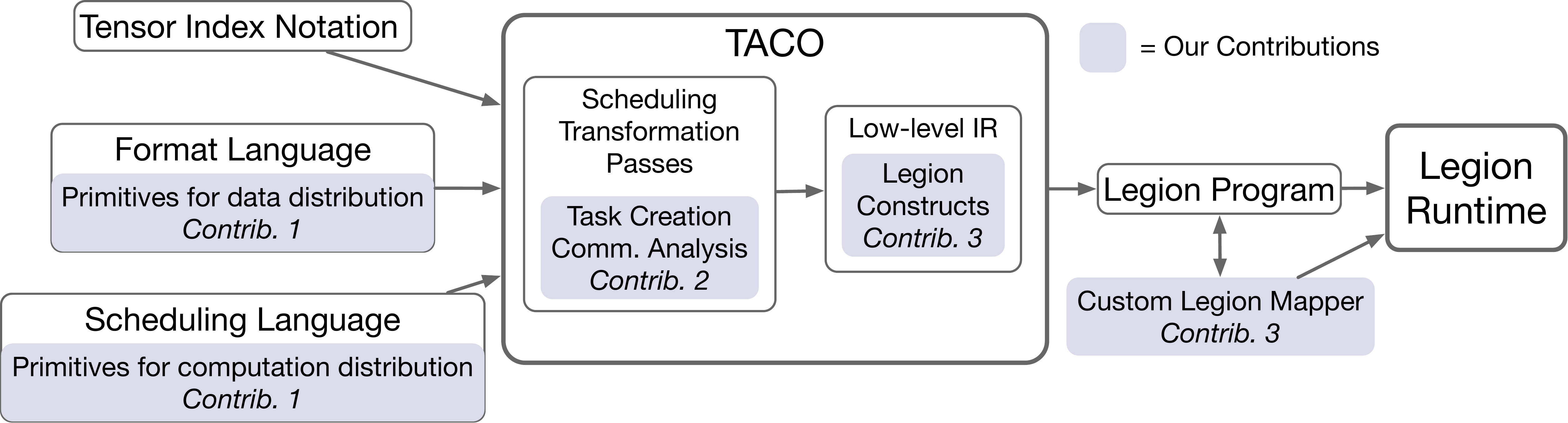}
    \caption{Overview of \name{}'s implementation.}
    \label{fig:system-overview}
\end{figure}

We implement \name{} by extending the dense functionality in the TACO~\cite{taco}
compiler to target the Legion~\cite{legion} distributed runtime system,
as shown in \autoref{fig:system-overview}.
We extend the format and scheduling languages of TACO with primitives for distribution.
Next, we add analysis passes and intermediate representation constructs to TACO to generate Legion
programs that interface with a mapper that places data and computation onto memories and processors.

The specific contributions of this work are:
\begin{enumerate}[leftmargin=1.5em]
    \item A data distribution language and set of scheduling commands that can express a wide variety of distributed tensor computations.
    \item A compiler that combines the separate specifications for how data and computation map onto distributed machines.
    \item An implementation of \name{} that extends the TACO~\cite{taco} compiler to target the Legion~\cite{legion} runtime system.
\end{enumerate}

We evaluate our contributions along two different axes:

\vspace{0.1em}
\noindent \textbf{Generality.} We implement several tensor algebra kernels and show that we achieve good
    performance. We show that our approach of generating bespoke implementations achieves
    between a 1.8x to 3.7x (with a 45.7x outlier) speedup over CTF, a system aimed at a similar level of generality.

\vspace{0.1em}
\noindent \textbf{Absolute Performance.}
Our system matches or outperforms existing systems on dense matrix-matrix multiplication, a tensor algebra operation that has been extensively optimized in prior work.
In particular, dense matrix multiplication code generated by \name{} outperforms the CTF~\cite{ctf} and ScaLAPACK~\cite{scalapack} libraries by at least 1.25x and comes within 0.95x the performance of COSMA~\cite{cosma}, the best published dense matrix-matrix multiplication implementation.

\begin{figure}
  \centering
  \begin{lstlisting}[frame=single,language=C++,escapechar=|,commentstyle=\color{gray},numbers=left,keywords={Format,Tensor,IndexVar,int,Machine,Grid,Param,Distribution,double,MappingVar,divide,communicate,substitute,reorder,DistributedGPU,CuBLAS,GeMM,distribute,schedule,compile,Dense,split,Memory},label={lst:lang-sample},basicstyle=\ttfamily\scriptsize]
// Declare input parameters for generated code.|\label{fig:lang-sample:line:params}|
Param gx, gy, n, chunkSize;
// Define the target machine m as a 2D grid of processors.
Machine m(Grid(gx, gy));|\label{fig:lang-sample:line:machine}|

// A tensor's format describes how it is distributed onto m.
// The following format partitions the two dimensions of a tensor by
// the two dimensions of m, resulting in a two-dimensional tiling.
// The final argument declares the tensor should reside in GPU
// framebuffer memory for fast access from GPUs.
Distribution tiles(m, {0, 1}, Memory::GPU_MEM); // (x, y) -> m(x, y)|\label{fig:lang-sample:line:datadist}|
Format f({Dense, Dense}, tiles);

// Declare three dense matrices with the same format.
Tensor<double> A({n, n}, f), B({n, n}, f), C({n, n}, f);|\label{fig:lang-sample:line:dist-format}|

// Declare the computation, a matrix-matrix multiply.
IndexVar i, j, k;|\label{fig:lang-sample:line:tensor-index-notation-begin}|
A(i, j) = B(i, k) * C(k, j);|\label{fig:lang-sample:line:tensor-index-notation}|

// Map the computation onto m via scheduling commands.
IndexVar io, ii, jo, ji, ko, ki;
A.schedule()|\label{fig:lang-sample:line:sched-begin}|
 // Tile i and j for each GPU.
 .divide(i, io, ii, m.x).divide(j, jo, ji, m.y)|\label{fig:lang-sample:line:dist-start}|
 .reorder({io, jo, ii, ji})
 // Distribute each i and j tile over all GPUs.
 .distribute(io).distribute(jo)|\label{fig:lang-sample:line:dist-end}|
 // Break the k loop into chunks.
 .split(k, ko, ki, chunkSize)
 // Move the outer k loop to outside the ii and ji loops.
 .reorder({io, jo, ko, ii, ji, ki})
 // Choose the granularity at which communication occurs.
 // Here, each processor operates on a local piece of a, and
 // receives chunks of b and c as the ko loop steps.
 .communicate(a, jo).communicate({b, c}, ko)
 // Schedule at leaves for variables ii, ji, and ki. Our
 // system can generate code for GPUs, but allows for using
 // heavily optimized kernels when applicable.
 .substitute({ii, ji, ki}, CuBLAS::GeMM)|\label{fig:lang-sample:line:sched-end};

  \end{lstlisting}
  \caption{Multi-GPU matrix multiplication in \name{} implementing the SUMMA~\cite{summa} algorithm used by ScaLAPACK~\cite{scalapack}.}
  \label{fig:lang-sample}
\end{figure}

\section{Background}
\label{sec:background}
\ignore{
\begin{figure}
    \small
    \[
        \begin{array}{rlcl} \\
        \textit{Index Variables} & i & & \\
        \textit{Numeric Constants} & c & & \\
        \textit{Tensors} & \mT & & \\
        \textit{Accesses} & a & \bnfdef & \mT(i^+) \\
        \textit{Expressions} & e & \bnfdef & a \bnfalt c \bnfalt e + e \bnfalt e * e \\
        \textit{Statements} & s & \bnfdef & a = e \\
        \end{array}
    \]
    \caption{Syntax of Tensor Index Notation \TODO{rohany: I think that this could be cut?}}
    \label{fig:tensor-index-notation-syntax}
\end{figure}
}

\autoref{fig:lang-sample} shows \name{}'s three input sub-languages:
a {\em computation language} that describes the desired kernel (lines \ref{fig:lang-sample:line:tensor-index-notation-begin}--\ref{fig:lang-sample:line:tensor-index-notation}), a {\em scheduling language} that describes how to optimize
the computation (lines \ref{fig:lang-sample:line:sched-begin}--\ref{fig:lang-sample:line:sched-end}), and a {\em format language} that describes how the tensors are stored (lines \ref{fig:lang-sample:line:machine}--\ref{fig:lang-sample:line:dist-format}).
In this section, we give background on each of these three components.

Computation is described in \name{} using {\em tensor index notation}, a domain specific language
used as input to the TACO compiler~\cite{taco}.
Tensor index notation consists of \emph{accesses} that index tensor dimensions with lists of variables.
Tensor index notation statements are assignments, where the left-hand side is an access, and the right-hand side
is an expression constructed from the addition and multiplication of accesses.
For example, the tensor-times-vector operation is expressed in tensor index notation as $A(i, j) = \sum_k B(i, j, k) \cdot c(k)$.
Each component $A(i, j)$ is the result of the inner-product
of the last dimension of $B$ with $c$.
Index variables correspond to nested loops, and variables used only on the right-hand side
represent sum reductions over their domain.
Tensor index notation allows for any number of tensors on the right-hand side
of an expression, and, like TACO, \name{} can generate a single fused kernel for the entire
tensor index notation expression.

The computation description is separated from the exact
algorithm to perform the computation through a scheduling
language~\cite{halide, graphit, TVM, tensorcomprehensions, tiramisu, taco_scheduling}.
We provide the following transformations introduced by prior systems~\cite{halide,taco_scheduling,taco_workspaces}:
\begin{itemize}
    \item \lstinline!parallelize!: parallelize the iterations of a loop
    \item \lstinline!split!/\lstinline!divide!: break a loop into an inner and outer loop
    \item \lstinline!collapse!: fuse two nested loops into a single loop
    \item \lstinline!reorder!: switch the execution order of two loops
    \item \lstinline!precompute!: hoist the computation of a subexpression %
\end{itemize}
We will introduce three new scheduling commands that describe how computations map onto a distributed machine.

Moreover, the TACO~\cite{taco} compiler introduced a format language that allows users to specify the
sparse format of each tensor in a computation.
While this work considers only dense computations, we take inspiration from TACO
to describe a tensor's distribution as part of the tensor's format.

\section{Core Abstractions}
\label{sec:abstractions}
This section describes the core abstractions of \name{} that allow
users to express a virtual machine organization and to map data and computation onto that machine.

\subsection{Modeling Modern Machines}

\textbf{}\name{} models a distributed machine $\mM$ as a multidimensional grid of abstract processors that each
have an associated local memory and can communicate with all other processors.
The purpose of the grid abstraction is twofold: to expose locality in the model (which may or may not exist in the physical machine)
and to match the grid-like structure of tensor algebra computations.

A flat machine representation is useful, but is not sufficient to model
many modern high performance systems.
These systems are often heterogeneous, where each node contains multiple accelerators and CPU sockets
that offer faster communication within a node than between nodes.\footnote{Inter-node communication is affected
by hierarchy in the network itself. For example, communication within a rack is faster than between racks.}
Therefore, our machine abstraction is also hierarchical: each
abstract processor may itself be viewed as a distributed machine.
We use this abstraction in our evaluation to model the Lassen supercomputer; we arrange
the nodes into multi-dimensional grids, then model each node as a grid of GPUs.

\begin{figure}[t]
    \footnotesize
    \[
        \begin{array}{rlrlrl}
            \textit{Machines} & \mM & \textit{Tensors} & \mT & \textit{Dimension Variables} & d\\
        \end{array}
    \]
    \vspace*{-1.5em}
    \[
        \begin{array}{rlcl}
        \textit{Dimension Name} & n & \bnfdef & d \bnfalt \mathbb{N} \bnfalt \text{'$*$'} \\
        \textit{Tensor Distribution} & \mD & \bnfdef & \tdist{d^+}{n^+} \\
        \end{array}
    \]
    \caption{Syntax of Tensor Distribution Notation}\label{fig:tensor-dist-syntax}
\end{figure}

\begin{figure*}
    \begin{subfigure}[b]{0.3\textwidth}
        \centering
        \includegraphics[width=0.8\textwidth]{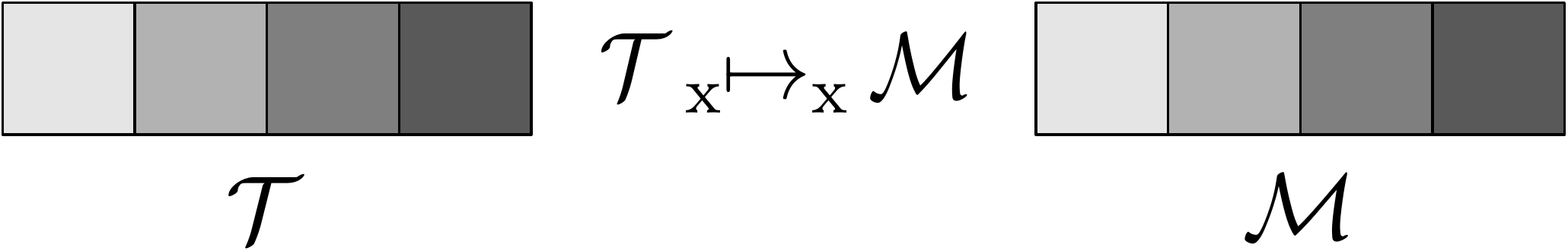}
        \captionof{figure}{Blocked distribution of a vector.}
        \label{fig:ti_mt_mi}
    \end{subfigure}\hfill
    \begin{subfigure}[b]{0.3\textwidth}
        \centering
        \includegraphics[width=0.8\textwidth]{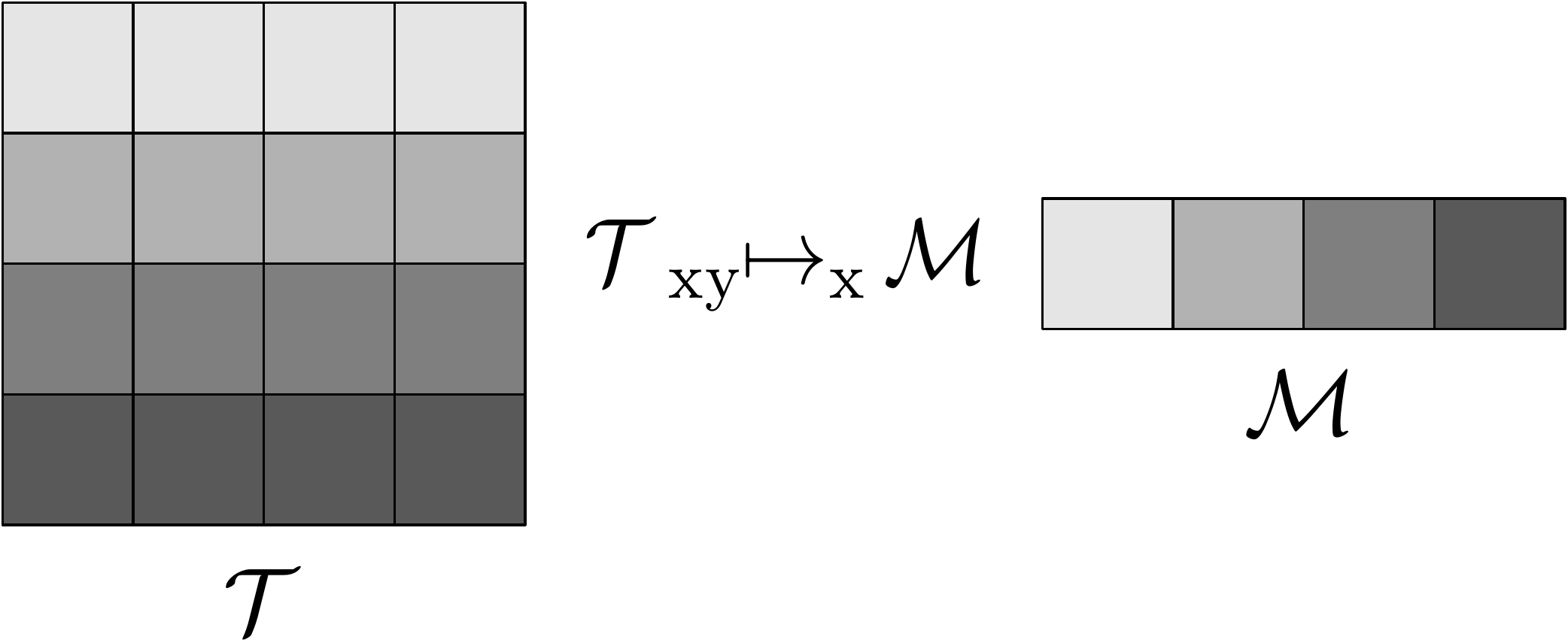}
        \captionof{figure}{Row-wise matrix distribution.}
        \label{fig:tij_mt_mi}
    \end{subfigure}\hfill
    \begin{subfigure}[b]{0.3\textwidth}
        \centering
        \includegraphics[width=0.8\textwidth]{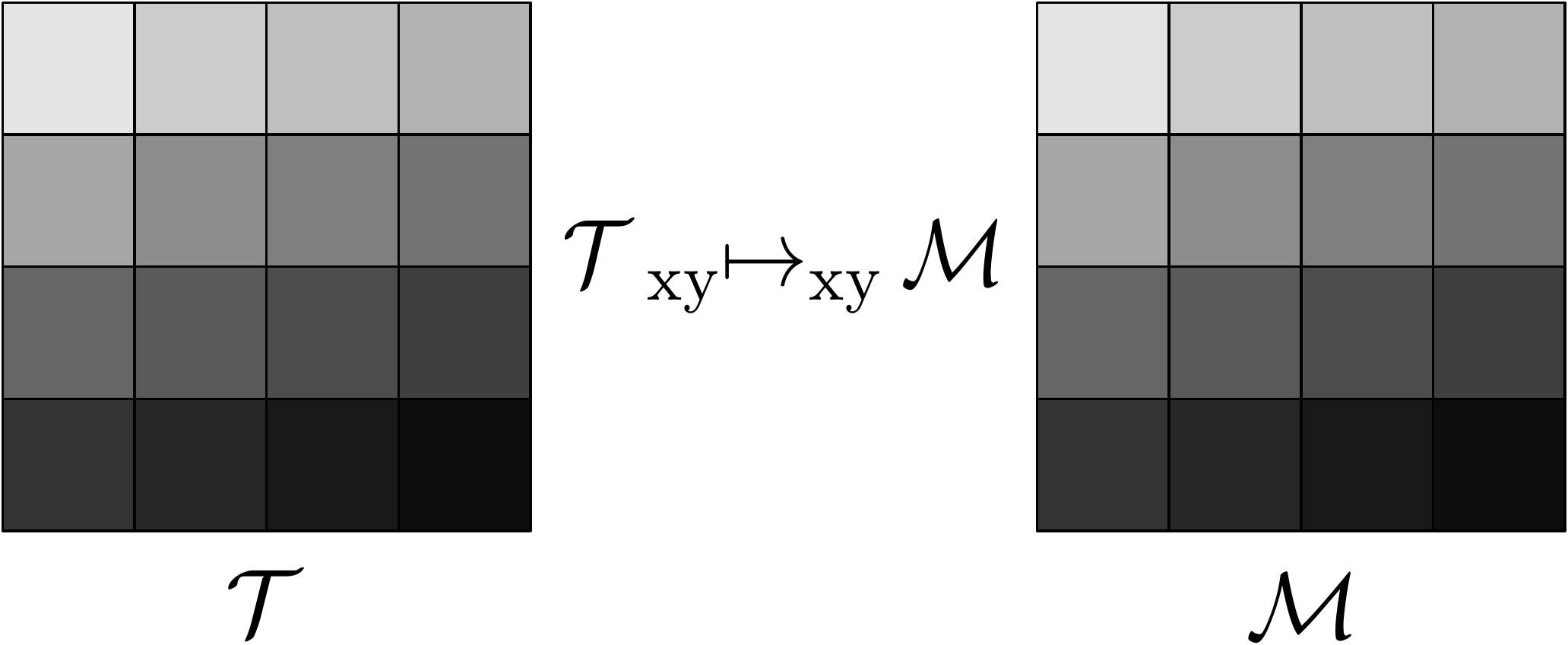}
        \captionof{figure}{Tiled matrix distribution.}
        \label{fig:tij_mt_mij}
    \end{subfigure}

    \vspace{0.5em}

    \begin{subfigure}[b]{0.3\textwidth}
        \centering
        \includegraphics[width=0.8\textwidth]{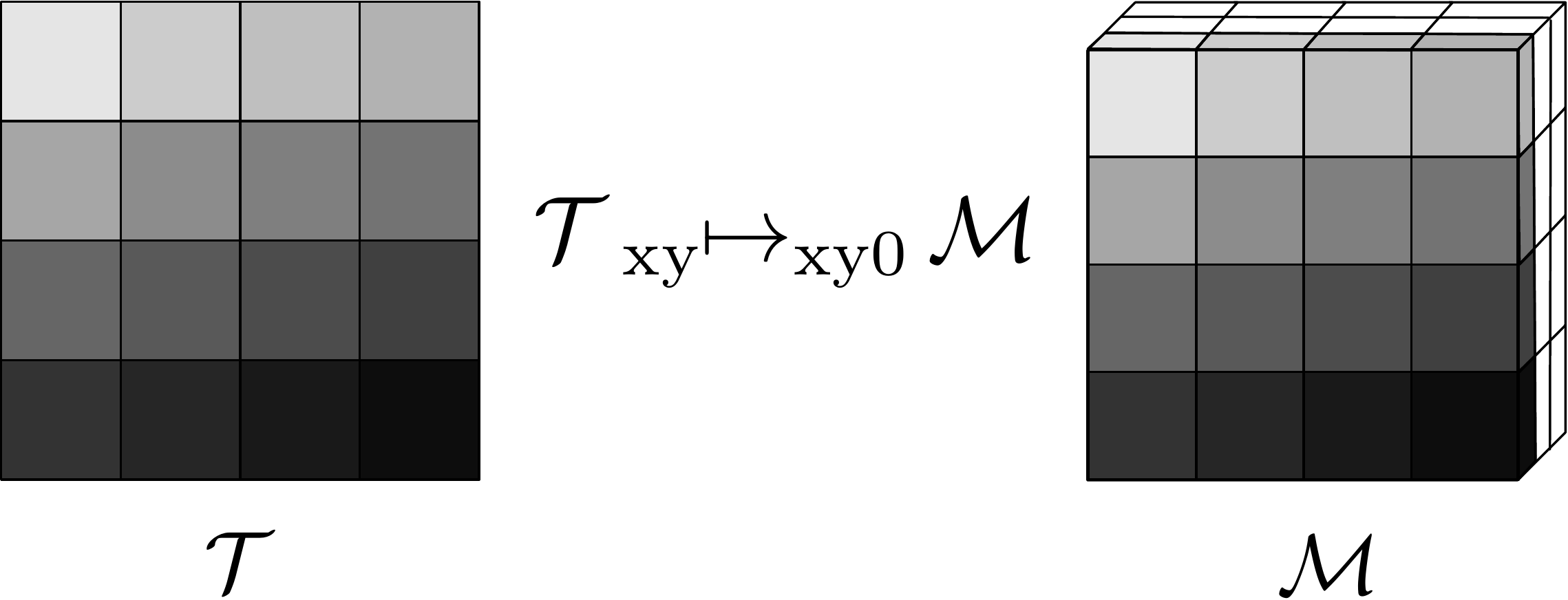}
        \caption{Fix a partition to a dimension.}
        \label{fig:tij_mt_mij0}
    \end{subfigure}\hfill
    \begin{subfigure}[b]{0.3\textwidth}
        \centering
        \includegraphics[width=0.8\textwidth]{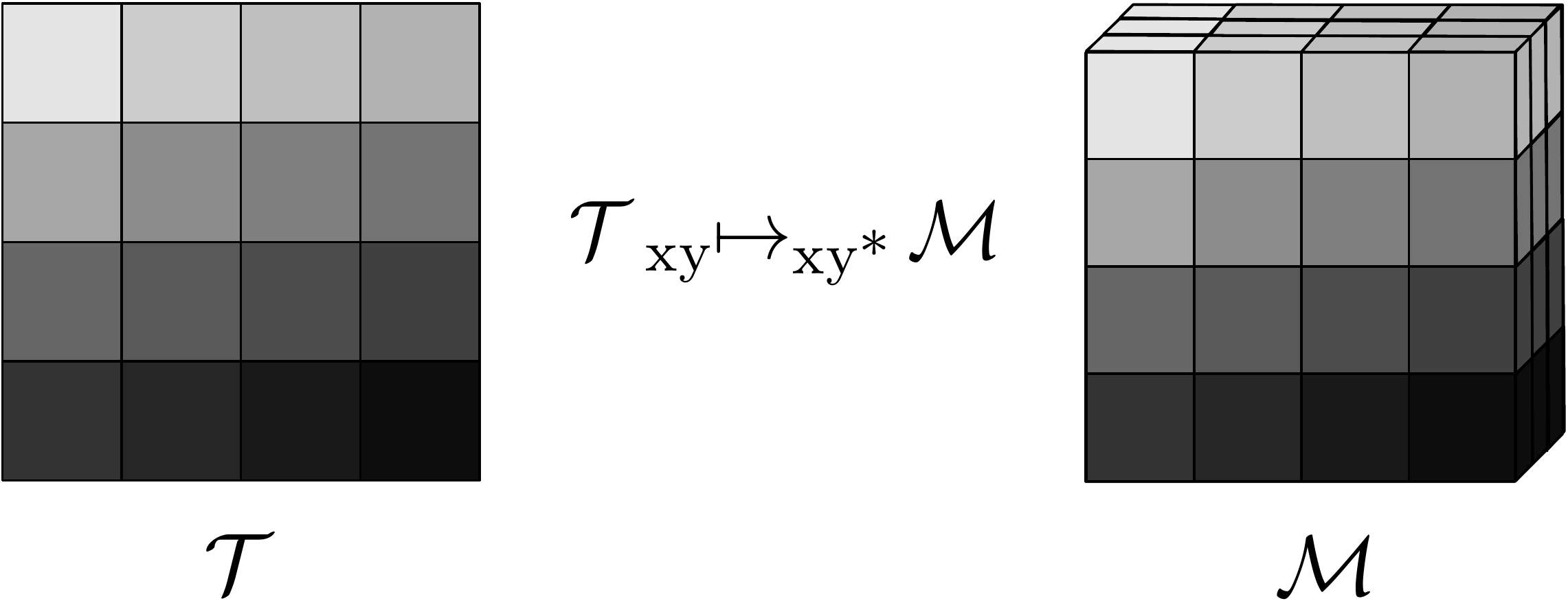}
        \caption{Broadcast a partition over a dimension.}
        \label{fig:tij_mt_mijstar}
    \end{subfigure}\hfill
    \begin{subfigure}[b]{0.3\textwidth}
        \centering
        \includegraphics[width=0.8\textwidth]{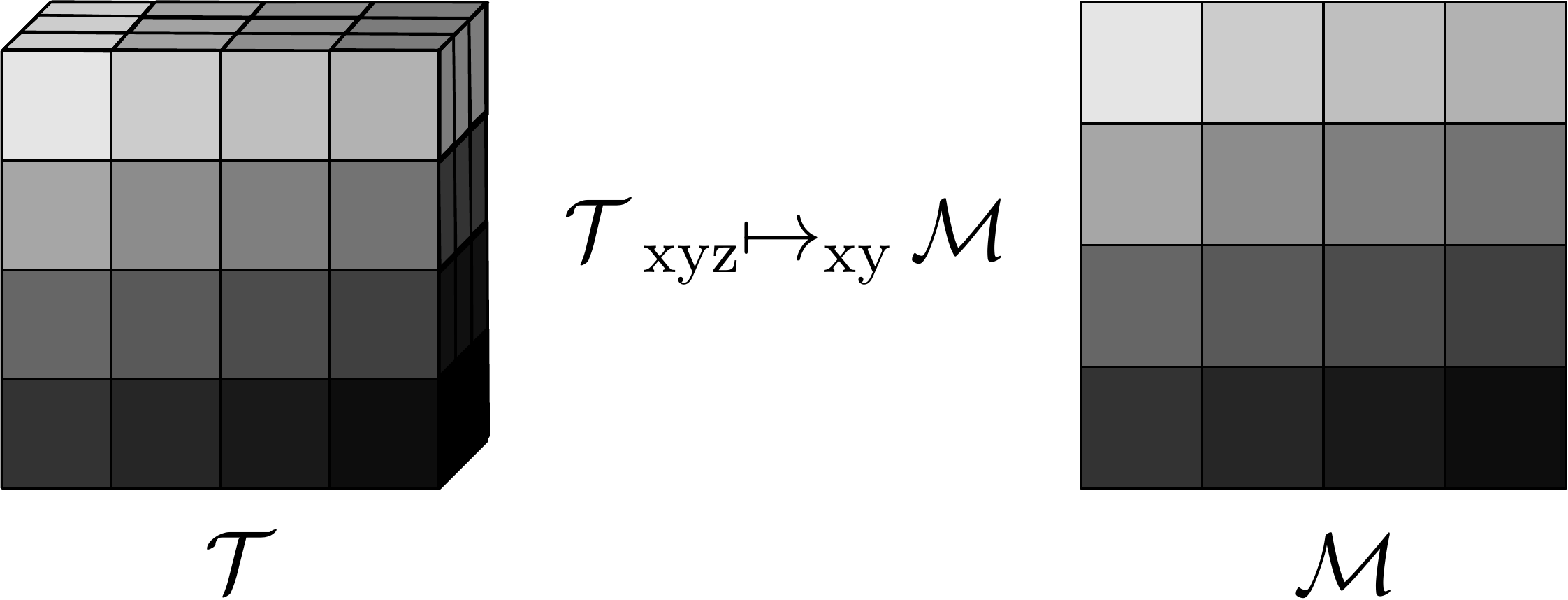}
        \caption{Map a 3-tensor onto a processor grid.}
        \label{fig:tijk_mt_mij}
    \end{subfigure}
    \caption{Examples of tensor distribution notation statements that map tensors onto machines in different ways.}
    \label{fig:example-data-distributions}
\end{figure*}

\subsection{Data Distribution}

Users map tensors onto machines through a sub-language of the format language called {\em tensor distribution notation}.
Tensor distribution notation allows users to describe how input tensors are already distributed in their program
or to move data into a distributed layout that suits the computation.

\textbf{Syntax.}
\autoref{fig:tensor-dist-syntax} describes the syntax for tensor distribution notation, which encodes how
dimensions (modes) of a tensor $\mT$ map onto the dimensions of a machine $\mM$.
Tensor distribution notation expresses this mapping by naming each dimension of $\mT$ and $\mM$; index sequences
on the left and right of the $\mapsto$ name dimensions in $\mT$ and $\mM$ respectively.
Tensor dimensions that share names with machine dimensions are \emph{partitioned} across those machine dimensions.
The remaining machine dimensions \emph{broadcast} the partition over the dimensions, or \emph{fix} the
partition to a coordinate in the dimension.
A tensor distribution notation statement $\tdist{X}{Y}$, where X and Y are index sequences, is valid if $|X| = \dim \mT$, $|Y| = \dim \mM$, both $X$ and $Y$ contain no duplicate
names, and all names in $Y$ are present in $X$.

\textbf{Intuition.}
Tensor dimensions partitioned across machine dimensions are divided into equal-sized contiguous pieces.
For example, if $\mT$ and $\mM$ are one-dimensional with 100 components and 10 processors respectively, then
the distribution $\tdist{x}{x}$ maps 10 components of $\mT$ to every processor of $\mM$.
Many common blocked partitioning strategies can be expressed by the choice of tensor dimensions to partition.
\autoref{fig:example-data-distributions} displays multiple ways of partitioning a matrix,
such as by rows ($\tdist{xy}{x}$, \autoref{fig:tij_mt_mi}), by columns ($\tdist{xy}{y}$),
or by two-dimensional tiles ($\tdist{xy}{xy}$, \autoref{fig:tij_mt_mij}).
Tensor dimensions that are not partitioned span their full extent in each partitioned piece,
as seen in \autoref{fig:tij_mt_mi} and \autoref{fig:tijk_mt_mij}.

\sloppy Machine dimensions that do not partition tensor dimensions either fix the tensor to
a single index or broadcast the tensor across all indices of a dimension.
A tensor is fixed to a single machine index by naming the dimension with a constant as in $\tdist{xy}{xy0}$ in \autoref{fig:tij_mt_mij0}, which restricts the tensor tiles to a face of the machine.
Marking a dimension with a $*$ replicates a tensor across that dimension, such as $\tdist{xy}{xy*}$ in \autoref{fig:tij_mt_mijstar} where tensor tiles are replicated across the entire third dimension of $\mM$.

\textbf{Semantics.}
Having described tensor distribution notation intuitively, we now provide a more formal description.
We define a tensor $\mT$ to be a set of coordinate--value pairs and a machine
$\mM$ to be a set of coordinates.
A tensor distribution notation statement $\tdist{X}{Y}$ is a function that maps the coordinates in $\mT$
to a non-empty set of processor coordinates in $\mM$, where $X$ and $Y$ are index sequences.
This function is the composition of two functions $\mP: \mT \rightarrow \textsf{color}$
and $\mF: \textsf{color} \rightarrow \mM~\textsf{set}$.
$\mP$ is an abstract partitioning function that maps coordinates in $\mT$ to unique \emph{colors}.
$\mF$ then maps each color in the range of $\mP$ to processors in $\mM$.
That is, we first group coordinates of $\mT$ into equivalence classes (colors), and then map each equivalence
class to a processor (or processors, if broadcast) in $\mM$.

Let $p = X \cap Y$, the dimensions of $\mT$ that are partitioned by dimensions of $\mM$.
Concretely, a \emph{color} is a point in the $p \subseteq Y$ dimensions of $\mM$.
$\mP$'s coloring is a many-to-one mapping between points in the $p$ dimensions of $\mT$ and colors.
The coloring is lifted to the remaining non-partitioned dimensions of $\mT$ in the natural way: every coordinate $x \in \mT$ is
given the same color as $\mP(x_p)$, where $x_p$ is $x$ restricted to the $p$ dimensions of $\mT$.
We choose to use a blocked partitioning function that maps contiguous ranges of coordinates to the same color.
However, other functions such as a cyclic distribution that maps adjacent coordinates to different colors could also be used.
As a running example, consider the distribution $\tdist{xy}{xy*}$
(\autoref{fig:tij_mt_mijstar}), where $\mT$ is 2x2 and $\mM$ is 2x2x2.
For this tensor distribution notation statement, 
{\small
\begin{align*}
    \mP = \{(0, 0) \mapsto (0, 0), (0, 1) \mapsto (0, 1), (1, 0) \mapsto (1, 0), (1, 1) \mapsto (1, 1)\},
\end{align*}
}%
mapping the coordinates in the 2x2 matrix onto points in the first two dimensions of the 2x2x2 machine cube.

$\mF$ maps $\mP$'s coloring of $\mT$ to full coordinates of processors in $\mM$.
As discussed previously, each color is an assignment to the $p \subseteq Y$ dimensions of $\mM$
and can be mapped to a coordinate of a processor in $\mM$ by specifying an index for the remaining
$Y - p$ dimensions.
Since all names in $Y$ must be present in $X$ as a condition of tensor distribution validity, 
the remaining $Y - p$ dimensions must either fix or broadcast the partition.
$\mF$ expands the color to the remaining dimensions of $\mM$ by casing
on whether each dimension fixes or broadcasts the partition: fixed dimensions are
set to the target value and broadcasted dimensions are expanded to all possible
coordinates in the dimension.
In the running example,
{\small
\begin{align*}
    \mF = \{&(0, 0) \mapsto \{(0, 0, 0), (0, 0, 1)\}, (0, 1) \mapsto \{(0, 1, 0), (0, 1, 1)\},\\
            &(1, 0) \mapsto \{ (1, 0, 0), (1, 0, 1)\}, (1, 1) \mapsto \{(1, 1, 0), (1, 1, 1)\}\},
\end{align*}
}
expanding the coloring of $\mP$ to the third dimension of $\mM$.

\textbf{Hierarchy.} Data distributions can also be hierarchical to match the hierarchical structure of the machine.
If the target machine has a hierarchical structure, then a tensor distribution can be provided for each level in the machine.
For example, if the machine $\mM$ is organized as a 2-dimensional grid at the node level, and then a 1-dimensional
grid of GPUs within each node,
then the distribution
$[\tdist{xy}{xy}, \tdist{zw}{z}]$ represents a two dimensional tiling of a matrix at the
outer level, and row-wise partition of each tile for each GPU.

\subsection{Computation Distribution}
\label{sec:comp-dist}

Like a tensor distribution notation statement describes how a tensor is distributed across a machine, a schedule describes how the iteration space of an expression is transformed and distributed.
In this section, we introduce three new scheduling transformations on top of those introduced in the background section: {\em distribute}, {\em communicate}, and {\em rotate}.
The first two were used in \autoref{fig:lang-sample}.
Throughout this section, we use the computation $a(i) = \sum_j b(j)$ as a running example, which has loop
structure $\forall_i\forall_j~a(i) \pluseq b(j)$.
This computation sets each index of $a$ to be the sum of all elements in $b$.\footnote{The optimal way to compute this expression is to aggregate $b$ into a scalar and assign the scalar to each index in $a$. For the sake of a simple example, we present alternate (albeit arithmetically inefficient) schedules to illustrate key concepts of \name{}.}
In our examples, we consider a one-dimensional machine $\mM$, where $\tdistname{A}{x}{x}$, $\tdistname{B}{x}{x}$
and $|A| = |B| = |\mM| = 3$.

\textbf{Iteration Spaces.} The iteration space of loops in a
tensor algebra expression is a hyper-rectangular grid of points formed by
taking the Cartesian product of the iteration domain of each index variable in the input expression.
Each point in the iteration space represents a scalar operation that is the atomic unit of computation
in our model.

\textbf{Execution Spaces.}
We model the execution of an iteration space through an \emph{execution space}, which describes
when and where each iteration space point is executed.
An execution space has a processor dimension and a time dimension, describing what processor
an iteration space point executes on and at what relative time the point executes.
A mapping of iteration space points onto an execution space describes an execution strategy for the iteration space.
Each processor and point in time may be assigned one point $p$ in the iteration space as well
as communication operations to fetch tensor data needed by $p$ that logically occur before $p$; we discuss communication operations below.
An iteration space's default execution space mapping linearizes all points in time according to the ordering of
the iteration space dimensions and maps all points to the same processor.

\begin{figure}
    \includegraphics[width=0.42\linewidth]{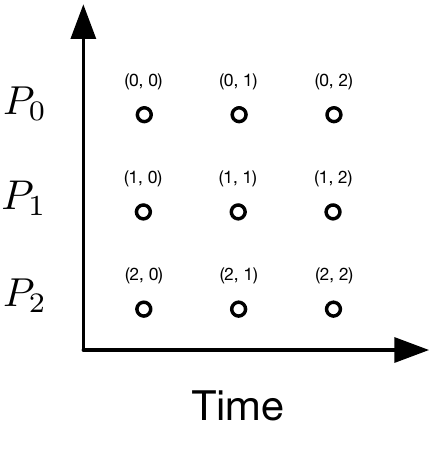}
    \caption{Execution space mapping of $\forall_i\forall_j~a(i) \pluseq b(j)$ after \lstinline!distribute(i)!. Iteration space points are labeled $(i, j)$.}
    \label{fig:exec-space-dist}
\end{figure}

\textbf{Distribute.} The \lstinline!distribute! operation transforms how the iteration space maps onto the execution space.
In particular, \lstinline!distribute!-ing a set of index variables $v$ modifies the execution space mapping
such that all iterations of the dimensions corresponding to $v$ occur on a different processors at the same time,
as seen in \autoref{fig:exec-space-dist}.\footnote{
The distribution of an iteration space $\mI$ onto a machine $\mM$ can also be viewed in a similar
manner to tensor distribution notation, where the desired $X \cap Y$ dimensions of $\mI$ are mapped
onto $\mM$ using $\tdistname{I}{X}{Y}$.
}

\ignore{
\TODO{rohany: I think that this paragraph doesn't add very much.}
The execution space mapping of iteration space points to processors in a machine $\mM$ can be viewed in a similar manner to
tensor distributions.
Using a similar notation, an iteration space $\mI$ is mapped onto $\mM$ through $\tdistname{\mI}{X}{Y}$
where $X \cap Y$ is the set of iteration space dimensions to distribute.
In this view, the iteration space coordinates are distributed onto processors in $\mM$ in the same way as
a tensor distribution maps coordinates in a tensor onto $\mM$.
}

The following compound command utilizes the \lstinline!distribute!, \lstinline!divide! and \lstinline!reorder! 
commands to map a set of iteration space dimensions onto a machine by tiling the iteration space
dimensions onto each processor in the machine:
\begin{lstlisting}[breaklines=false,numbers=left,basicstyle=\ttfamily\scriptsize]
distribute(vector<IndexVar> targets, vector<IndexVar> dist,
            vector<IndexVar> local, Machine m):
  for i in range(0, m.dim):
    # Divide each dimension by the corresponding machine dimension.
    divide(targets[i], dist[i], local[i], m.dims[i])
  # Reorder loops so each outer divided variable is on the outside.
  reorder(dist + local)
  # Distribute all of the outer divided variables.
  distribute(dist)
\end{lstlisting}

The choice of variables to distribute affects the resulting communication patterns.
Distributing variables that index the output tensor pull input tensors towards
a stationary output tensor in an owner-computes paradigm.
Distributing variables used for reductions results in distributed reductions into the output, trading space usage
for increased parallelism.

To match hierarchical machine models and data distributions, \lstinline!distribute! may also be applied hierarchically.
For example, we can apply this strategy in computations that benefit from locality, like matrix-multiply, to use a distributed algorithm
at the node level and another (sometimes different) algorithm for the multiple GPUs within a node.

\begin{figure}
    \begin{subfigure}{0.23\textwidth}
        \centering
        \includegraphics[width=\textwidth]{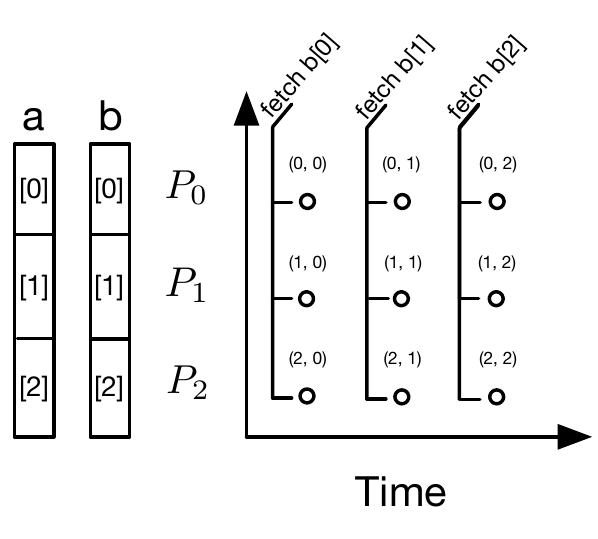}
        \caption{Na\"ive completion where communication is inserted as needed at each iteration space point.}
        \label{fig:exec-space-completed}
    \end{subfigure}\hfill
    \begin{subfigure}{0.23\textwidth}
        \centering
        \includegraphics[width=\textwidth]{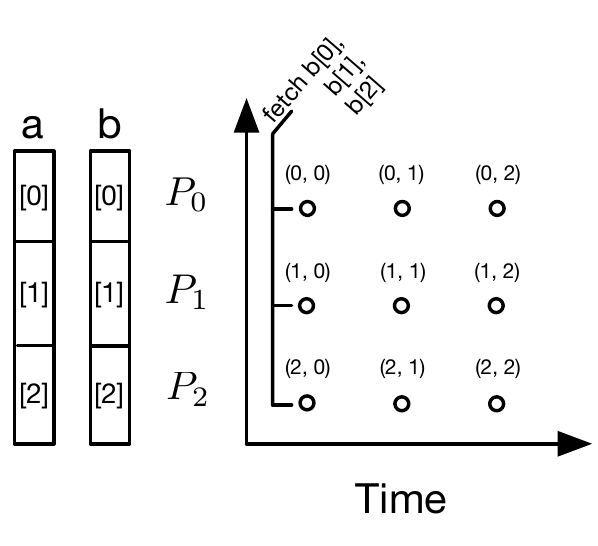}
        \caption{Completion where communication is aggregated underneath each $i$ iteration.}
        \label{fig:exec-space-comm}
    \end{subfigure}
    \caption{Completions of the distributed execution space mapping of
        $\forall_i\forall_j~a(i) \pluseq b(j)$ where $\tdistname{a}{x}{x}$ and $\tdistname{b}{x}{x}$.}
    \label{fig:exec-space-completions}
\end{figure}

\textbf{Communicate.}
Every iteration space point maps to a coordinate in each of the input
and output tensors, corresponding to the data required at that point.
The data required at each iteration space point may not be present in the memory of the processor the point is mapped to,
and must be communicated from a memory where the data resides.
A communication operation will be automatically inserted at each execution space point where the required data
is not present, resulting in a na\"ive \emph{completion} of the execution space mapping.
A na\"ive completion of the execution space of $\forall_i\forall_j~a(i) \pluseq b(j)$ transformed by \lstinline!distribute(i)!
is depicted in \autoref{fig:exec-space-completed}.

Communication operations can be made more efficient by aggregating them into larger operations that fetch
the data for a group of iteration space points, as seen in \autoref{fig:exec-space-comm}.
The choice of how much communication to aggregate incurs the tradeoff between memory usage and communication
frequency---more frequent communication allows for lower memory usage at the cost of more messages sent.

To allow for optimization over this tradeoff space, the \lstinline!communicate! command controls how much
communication for each tensor should be aggregated into a single message.
Precisely, \lstinline[mathescape=true]{communicate($\mT$, i)} aggregates the communication of $\mT$ at the beginning
of each iteration of \lstinline!i! by materializing the data for all iteration space points nested under each iteration of \lstinline!i!
in the executing processor's memory. If no \lstinline!communicate! command is given, then communication will be nested under the inner-most index variable.

It is a deliberate choice to omit any notion of processor ranks, explicit sends/receives, and specific channels used
(such as in prior work~\cite{tiramisu}) from the scheduling language so that schedules only affect performance,
not correctness, as well as to keep the scheduling language relatively simple.
The \lstinline!communicate! command is not needed for correctness and is used only
to optimize the communication pattern of the computation.

\begin{figure}
    \begin{subfigure}{0.23\textwidth}
        \centering
        \includegraphics[width=\textwidth]{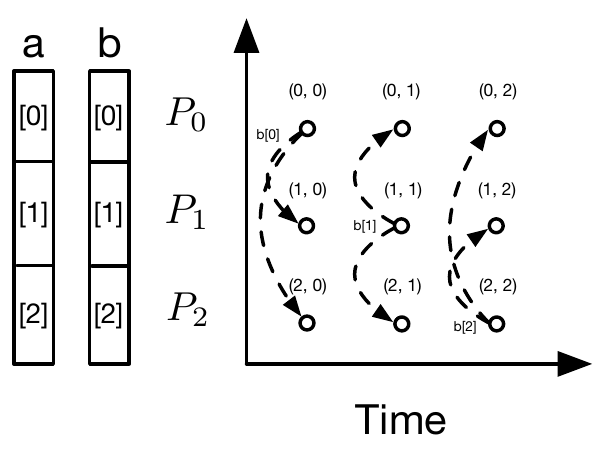}
        \caption{Standard execution space mapping. At each time step, the processor $j$ broadcasts $b[j]$ to
        all other processors.}
        \label{fig:exec-space-bcast}
    \end{subfigure}\hfill
    \begin{subfigure}{0.23\textwidth}
        \centering
        \includegraphics[width=\textwidth]{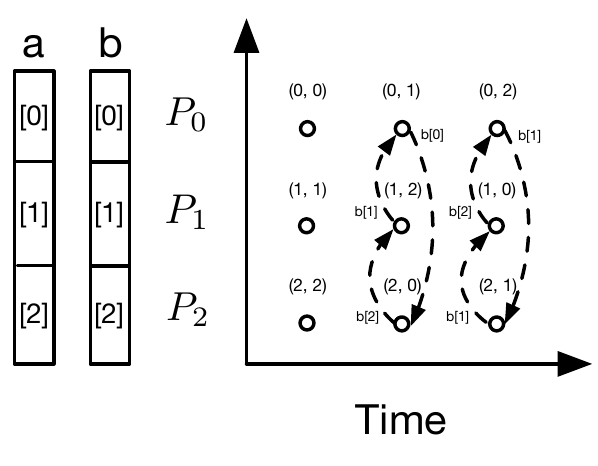}
        \caption{Rotated execution space mapping. At each time step, each processor transfers the data
        received at the previous time step.}
        \label{fig:exec-space-systolic}
    \end{subfigure}
    \caption{Communication (denoted by dashed arrows) between processors in execution space mappings for $\forall_i\forall_j~a(i) \pluseq b(j)$ with and without \lstinline!rotate!.}
    \label{fig:comm-partners}
\end{figure}

\renewcommand\tabularxcolumn[1]{m{#1}}%
\newcolumntype{Y}{>{\Centering}X}
\begin{figure*}[t]
    \small
    \begin{tabularx}{\textwidth}{|Y|Y|Y|Y|l|}
        \hline
        Algorithm & Comm. Pattern & Target Machine & Data Distribution & Schedule \\
        \hline
        Cannon's~\cite{cannon} \newline (1969) & \vspace{0.5em}\includegraphics[width=0.7in]{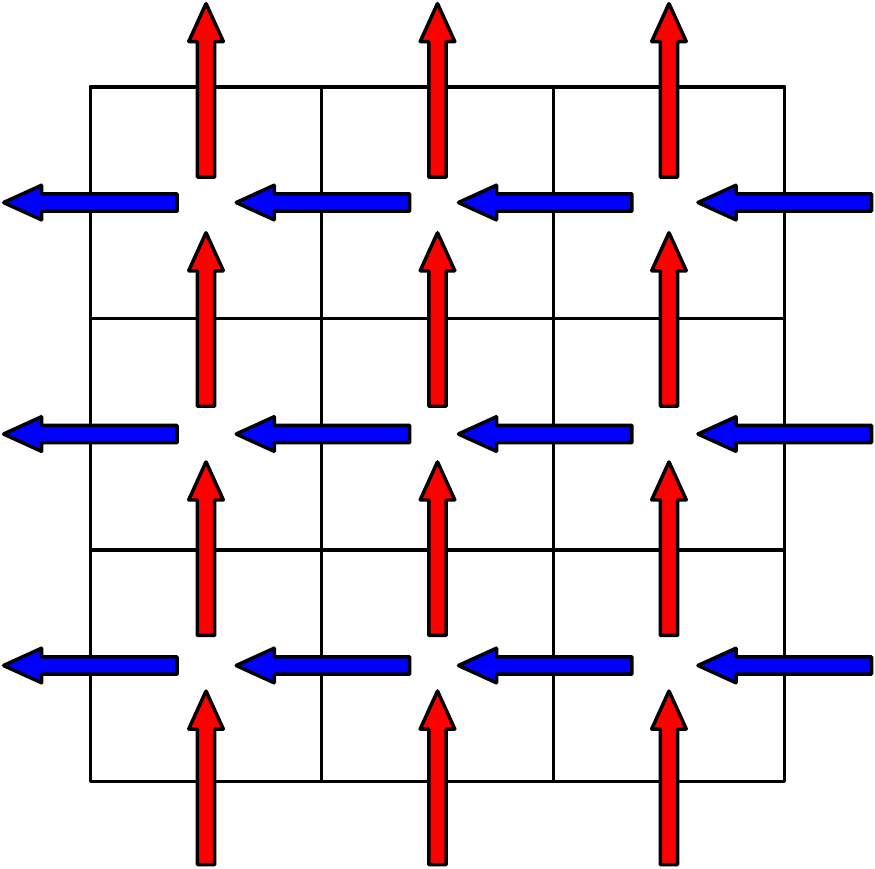} & $\mM = \textsf{Grid}(gx, gy)$ &
        $\tdistname{A}{xy}{xy}$ \newline $\tdistname{B}{xy}{xy}$ \newline $\tdistname{C}{xy}{xy}$
        & 
\begin{lstlisting}[basicstyle=\ttfamily\tiny]
.distribute({i, j}, {io, jo}, {ii, ji}, Grid(gx, gy))
.divide(k, ko, ki, gx)
.reorder({ko, ii, ji, ki})
.rotate(ko, {io, jo}, kos)
.communicate(A, jo).communicate({B, C}, kos)
\end{lstlisting} \\
        \hline
        PUMMA~\cite{pumma} \newline (1994) & \vspace{0.5em}\includegraphics[width=0.70in]{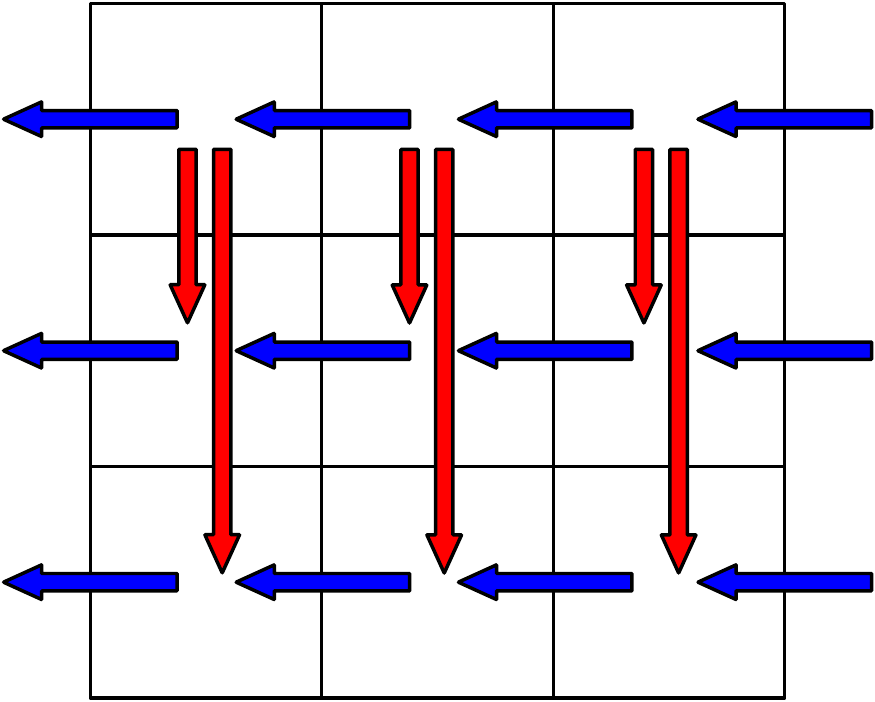} & $\mM = \textsf{Grid}(gx, gy)$ &
        $\tdistname{A}{xy}{xy}$ \newline $\tdistname{B}{xy}{xy}$ \newline $\tdistname{C}{xy}{xy}$
        & 
\begin{lstlisting}[basicstyle=\ttfamily\tiny]
.distribute({i, j}, {io, jo}, {ii, ji}, Grid(gx, gy))
.divide(k, ko, ki, gx)
.reorder({ko, ii, ji, ki})
.rotate(ko, {io}, kos)
.communicate(A, jo).communicate({B, C}, kos)
\end{lstlisting} \\
        \hline
        SUMMA~\cite{summa} \newline (1995) & \vspace{0.5em}\includegraphics[width=0.60in]{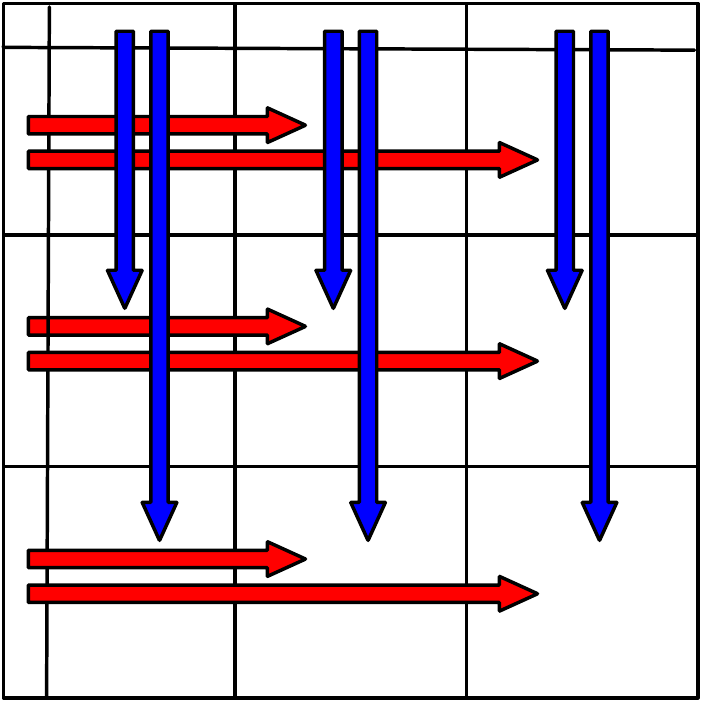} & $\mM = \textsf{Grid}(gx, gy)$ &
        $\tdistname{A}{xy}{xy}$ \newline $\tdistname{B}{xy}{xy}$ \newline $\tdistname{C}{xy}{xy}$
        &  
\begin{lstlisting}[basicstyle=\ttfamily\tiny]
.distribute({i, j}, {io, jo}, {ii, ji}, Grid(gx, gy))
.split(k, ko, ki, chunkSize)
.reorder({ko, ii, ji, ki})
.communicate(A, jo).communicate({B, C}, ko)
\end{lstlisting} \\
        \hline
        Johnson's~\cite{johnson} \newline (1995) & \vspace{0.5em}\includegraphics[width=0.60in]{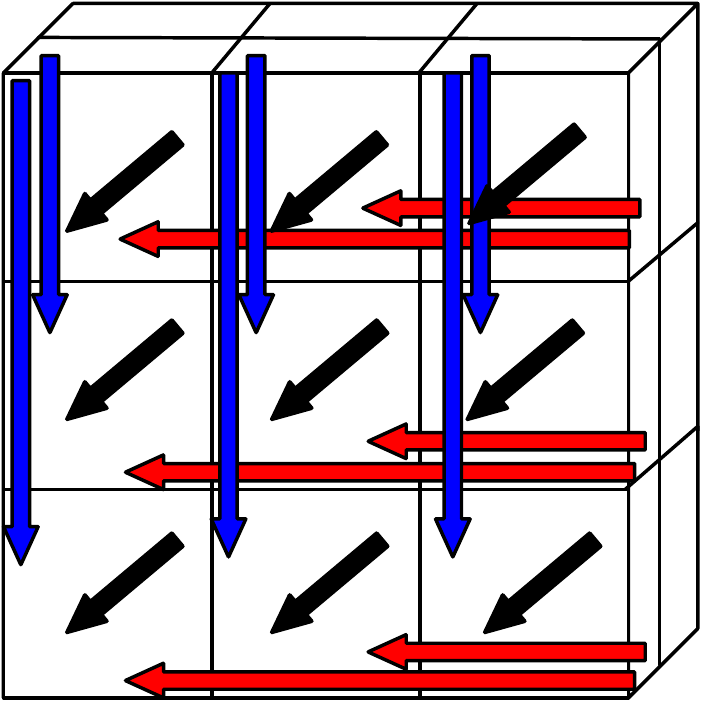} & $\mM = \textsf{Grid}(\sqrt[3]{p}, \sqrt[3]{p}, \sqrt[3]{p})$ &
        $\tdistname{A}{xy}{xy0}$ \newline $\tdistname{B}{xz}{x0z}$ \newline $\tdistname{C}{zy}{0yz}$
        & 
\begin{lstlisting}[basicstyle=\ttfamily\tiny,mathescape=true]
.distribute({i, j, k}, {io, jo, ko},
            {ii, ji, ki}, Grid($\sqrt[3]{p}$, $\sqrt[3]{p}$, $\sqrt[3]{p}$))
.communicate({A, B, C}, kn)
\end{lstlisting} \\
        \hline
        Solomonik's~\cite{solomonikMM} \newline (2011) & \vspace{0.5em}\hspace*{-0.4em}\includegraphics[width=0.65in]{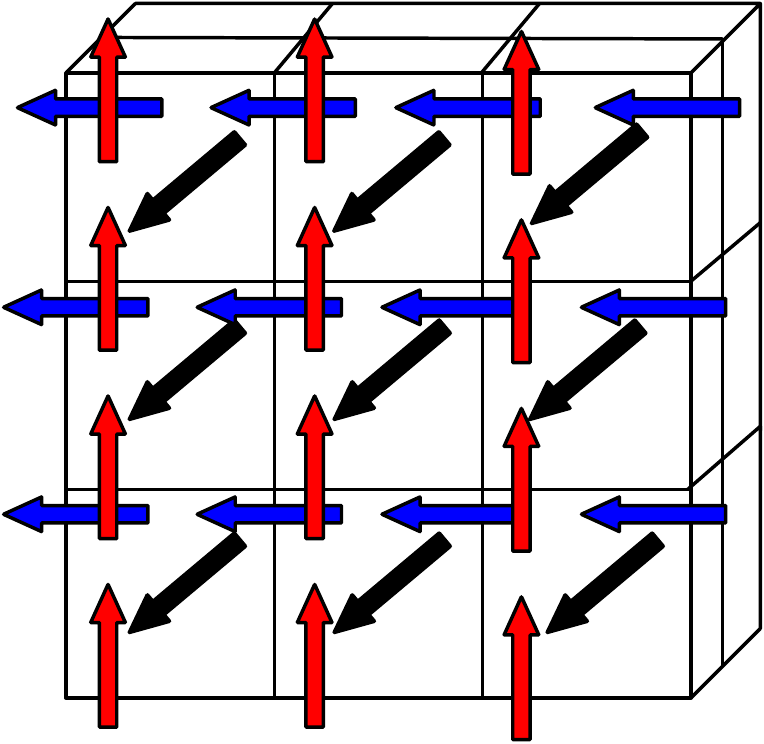} & $\mM = \textsf{Grid}(\sqrt{\frac{p}{c}}, \sqrt{\frac{p}{c}}, c)$ &
        $\tdistname{A}{xy}{xy0}$ \newline $\tdistname{B}{xy}{xy0}$ \newline $\tdistname{C}{xy}{xy0}$
        & 
\begin{lstlisting}[basicstyle=\ttfamily\tiny,mathescape=true]
.distribute({i, j, k}, {io, jo, ko},
            {ii, ji, ki}, Grid($\sqrt{\frac{p}{c}}$, $\sqrt{\frac{p}{c}}$, c))
.divide(ki, kio, kii, $\sqrt{\frac{p}{c^3}}$)
.reorder({kio, il, jl, kii})
.rotate(kio, {io, jo}, kios)
.communicate(A, jo).communicate({B, C}, kios)
\end{lstlisting} \\
        \hline
        COSMA~\cite{cosma} \newline (2019) & \vspace{0.5em}\includegraphics[width=0.60in]{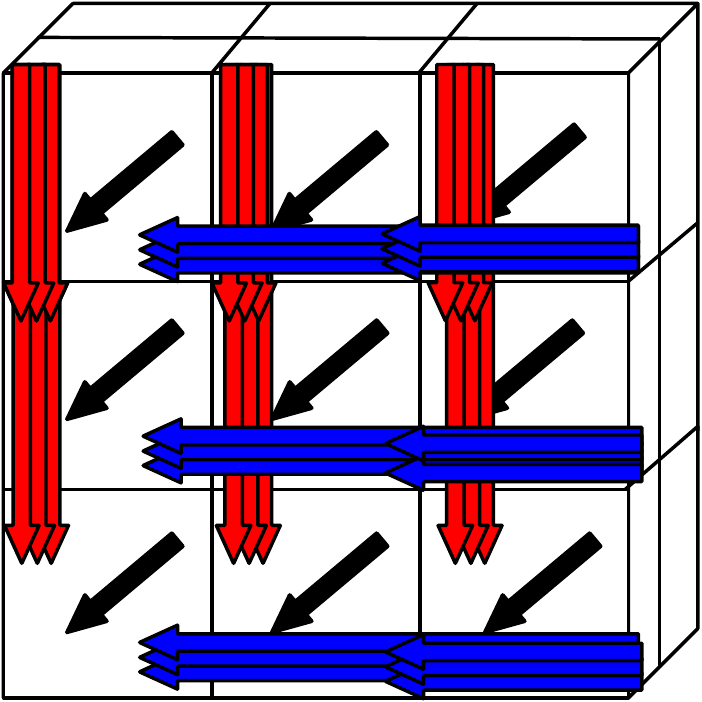} & induced by schedule & induced by schedule & 
\begin{lstlisting}[basicstyle=\ttfamily\tiny,mathescape=true]
// gx, gy, gz, numSteps computed by COSMA scheduler.
.distribute({i, j, k}, {io, jo, ko}
            {ii, ji, ki}, Grid(gx, gy, gz))
.divide(ki, kio, kii, numSteps)
.reorder({kio, il, jl, kii})
.communicate(A, ko).communicate({B, C}, kio)
\end{lstlisting}
        \\
        \hline
    \end{tabularx}
    \caption{Set of matrix-multiplication algorithms representable by \name{}. For each algorithm, we show the high level
    communication pattern, target machine organization, initial data distributions, and schedule of the compute statement
    $A(i, j) = \sum_k B(i, k) \cdot C(k, j)$. In the icons, black arrows indicate communications for $A$, blue for $B$ and red for $C$.
    The schedules utilize the compound \lstinline!distribute! command introduced in \autoref{sec:comp-dist}.
    }
    \label{fig:algorithm-schedules}
\end{figure*}

\textbf{Rotate.}
Many distributed algorithms have a systolic communication pattern, where processors repeatedly shift
data to their neighbors.
Systolic algorithms can take advantage of machine architectures with interconnects
that offer higher performance for nearest-neighbor communication and improve performance
by avoiding contention for the same pieces of data.
To express systolic computations, we introduce the \lstinline!rotate! operation, which
acts as a symmetry-breaking operation for distributed loops by transforming the mapping onto the
time dimension of the execution space.

To gain intuition for \lstinline!rotate!, consider the execution space mapping of the running
example $\forall_i\forall_j~a(i) \pluseq b(j)$ transformed by \lstinline!distribute(i)!.
At each time step, all processors access (and issue communication requests for) the same element
of $B$, causing the owner of that element to broadcast it to all other processors, as seen in
\autoref{fig:exec-space-bcast}.
In contrast, a systolic version of this computation would instead have each processor accumulate the local
element of $b$, and then shift the local element to the processor to its left in $\mM$.

The systolic communication pattern is achieved by changing the point in time that each processor
executes each iteration of the $j$ loop.
If the mapping of the time dimension is reordered such that for each iteration of $i$, the $j$ iteration space is
\emph{rotated} in time so that the $i$th iteration of $j$ occurs first, then the resulting execution space
has a systolic communication pattern, where each processor utilizes the data that the processor to
its right used in the previous iteration.
This effect is depicted in \autoref{fig:exec-space-systolic} where no processors execute the same iteration of
$j$ at the same point in time.

Concretely, given a set of index variables $I$, target index variable $t$ and result index variable $r$,
\lstinline[mathescape=true]{rotate($t$, $I$, $r$)} rotates each iteration of $t$ by $\sum_{i \in I} i~\textsf{mod extent}(t)$.
The effect of \lstinline!rotate! is that $\forall i \in I$, given a fixed iteration for all
remaining $i' \in (I - i)$, the same iteration of $r$ occurs at a different time
for all iterations of $i$.
For example, if $t$ is rotated by variables $i$ and $j$ oriented in a 2D grid,
every row starts the iteration of $r$ at a unique value, and vice versa
for each column.

\section{Matrix-Multiplication Case Studies}
\label{sec:algorithms}
Tensor distribution notation and scheduling can be composed to express a wide variety of algorithms.
A large body of research on distributed tensor algebra focuses on algorithms for distributed
matrix-multiplication.
Therefore, to showcase the expressivity of our techniques, we perform a case study on
matrix-multiplication algorithms discussed in the literature.
However, our techniques are not specific to matrix-multiplication and generalize to all of tensor algebra.

\subsection{Distributed Matrix-Multiplication Background}

The first distributed matrix-multiplication algorithm, presented by Cannon~\cite{cannon},
uses a systolic communication pattern and a tiled distribution of the input matrices.
The PUMMA~\cite{pumma} and SUMMA~\cite{summa} algorithms extended this work by generalizing to
rectangular matrices and improving the communication patterns through pipelining.
The SUMMA algorithm is implemented in the widespread ScaLAPACK~\cite{scalapack} library.
These are called {\em 2D algorithms}, because they organize the target machine into
a 2D grid and decompose the input matrices in tiles on the processor grid.

Follow up work by Agrawal et al.~\cite{johnson} introduced Johnson's algorithm, which organized
processors into a 3D grid and utilized extra memory per processor to perform asymptotically less communication
than 2D algorithms.
Algorithms of this style are called {\em 3D algorithms}.
The 2.5D algorithm by Solomonik et al.~\cite{solomonikMM} interpolates between 2D and 3D algorithms, utilizing
extra memory to reduce communication.
The 2.5D algorithm has been implemented in the Cyclops Tensor Framework~\cite{ctf}.
Finally, the COSMA~\cite{cosma} algorithm takes a different approach by computing an optimal processor
organization and parallelization strategy depending on the target matrix dimensions and machine size.

\autoref{fig:algorithm-schedules} depicts the communication pattern of these algorithms and demonstrates
how each can be implemented in \name{}.
Although \name{} cannot represent recursive algorithms like CARMA~\cite{carma}, it still covers a space of
widely used algorithms.
We now present detailed derivations of SUMMA, Cannon's and Johnson's algorithms.

\subsection{SUMMA}
MPI-like pseudocode for SUMMA can be found in \autoref{fig:summa-alg}.
SUMMA organizes the computation into a 2D grid and each processor owns a tile of $A$, $B$, and $C$.
Computation proceeds in chunks over the $k$ loop, where processors owning the $k$'th chunks
of the $B$ and $C$ loops broadcast the chunks within their row and column respectively.
Then, each processor multiplies the communicated chunks of $B$ and $C$ into a local tile of $A$.

\begin{figure}%
  \begin{lstlisting}[mathescape=tru,frame=single,commentstyle=\color{gray},language=none,basicstyle=\ttfamily\scriptsize]
# Arrange $p$ processors into a 2D grid.
# Assign a tile of $A$, $B$, $C$ to each processor.
for all $P_{ij}$ in parallel:
  for kc in (0, k, chunkSize):
    $B_l$ = row broadcast the kc to kc+chunkSize columns of $B$
    $C_l$ = col broadcast the kc to kc+chunkSize rows of $C$
    A += $B_l$ $\times$ $C_l$
  \end{lstlisting}
  \caption{Pseudocode for the SUMMA algorithm.}
  \label{fig:summa-alg}
\end{figure}

\sloppy SUMMA organizes the target machine as a 2D grid ($\mM = \textsf{Grid(gx, gy)}$)
and maps each tensor in tiles using $\tdistname{A}{xy}{xy}$, $\tdistname{B}{xy}{xy}$, and $\tdistname{C}{xy}{xy}$.
The $i$ and $j$ iteration space dimensions are distributed, and every processor locally iterates through
the $k$ dimension.
Therefore, we apply \lstinline!distribute({i,j}, {io,jo}, {ii,ji}, Grid(gx,gy))!.
Next, SUMMA steps through $k$ in chunks
expressible with \lstinline!split(k, ko, ki, chunkSize)! followed by \lstinline!reorder({ko, ii, ji, ki})!.
Finally, we schedule the communication: since each processor operates on a tile of $A$,
we \lstinline!communicate(A, jo)!.
SUMMA broadcasts $B$ and $C$ in chunks along the $k$ loop, so we communicate under
the \lstinline!ko! outer loop using \lstinline!communicate({B, C}, ko)!.

This schedule implements SUMMA.
Each processor steps over the $k$ dimension of the iteration space, and performs local matrix multiplications
on chunks of $B$ and $C$.
Every processor operates on the same chunks of $B$ and $C$ in its row and column, the processors that owns
the chunks broadcast them to the other processors in their rows and columns.

\subsection{Cannon's Algorithm}
\begin{figure}
  \includegraphics[width=\linewidth]{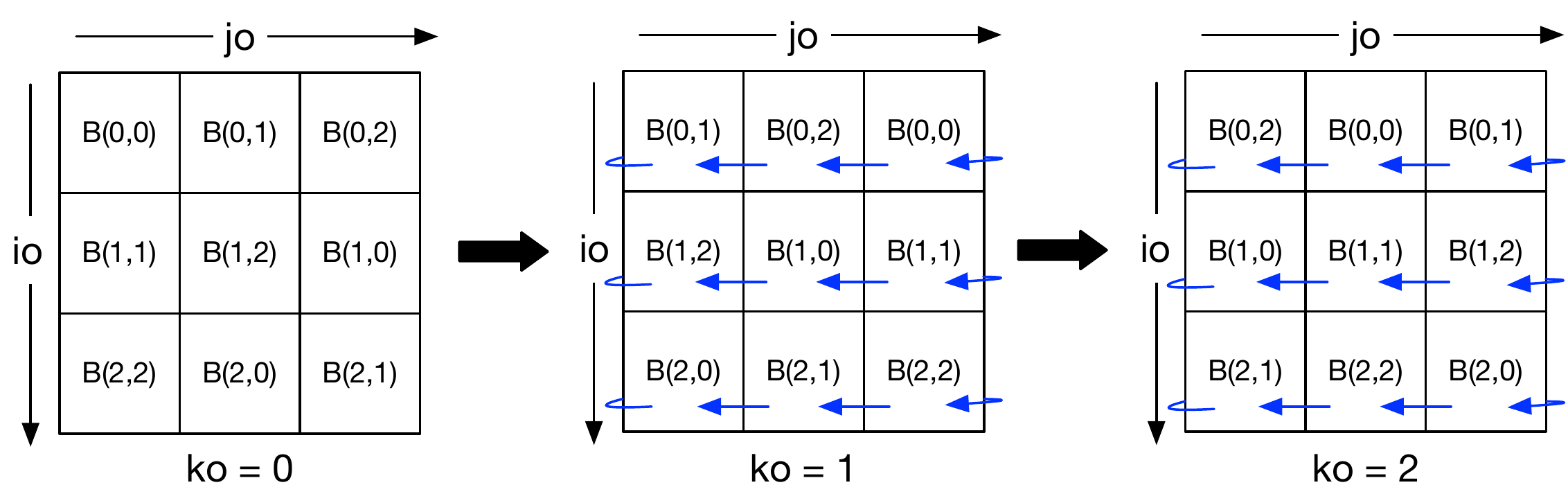}
  \caption{Communication pattern of $B$ in the Cannon's algorithm schedule on a 3x3 grid
  of processors.
  $B(x, y)$ is the $(x,y)$th tile of $B$.
  At each iteration of \lstinline!ko!, each processor
  performs the rotated iteration \lstinline[keepspaces]!kos = ko + io + jo mod 3!, accessing $B($\lstinline!io, kos!$)$.
  Each processor is labeled with the tile of $B$ needed at the current iteration.
  Blue arrows indicate from where the data needed at the current iteration was sent.}
  \label{fig:cannon-systolic}
\end{figure}

As another 2D algorithm, Cannon's algorithm has the same target machine and data distribution as SUMMA,
but has a systolic communication pattern, as seen in \autoref{fig:cannon-alg}, where processors shift
tiles along their row and column.
Despite the differences in pseudocode, the schedule for Cannon's algorithm is similar to the schedule for SUMMA's.
First, we change the \lstinline!split! operation into \lstinline!divide(k, ko, ki, gx)!
so that the tiles communicated are the same size as the tiles held by each processor.
Next, we change the communication pattern to the systolic pattern using \lstinline!rotate!.
We insert a \lstinline!rotate(ko, {io,jo}, kos)! to rotate each processors \lstinline!ko! iteration space
by the processor's coordinate in the processor grid.
\autoref{fig:cannon-systolic} depicts the communication pattern of $B$ after rotation.

\begin{figure}%
  \begin{lstlisting}[mathescape=tru,frame=single,commentstyle=\color{gray},language=none,basicstyle=\ttfamily\scriptsize]
# Arrange $p$ processors into a 2D grid, $\sqrt{p} \times \sqrt{p}$.
# Assign a tile of $A$, $B$, $C$ to each processor.
# Perform an initial data shift.
for all $P_{ij}$ in parallel:
  shift $B_{ij}$ $i$ spaces to the left
  shift $C_{ij}$ $j$ spaces upwards
for all $P_{ij}$ in parallel:
  for k in (0, $\sqrt{p}$):
    $A_{ij}$ += $B_{ij}$ $\times$ $C_{ij}$
    shift $B_{ij}$ to the left
    shift $C_{ij}$ upwards
  \end{lstlisting}
  \caption{Pseudocode for Cannon's algorithm.}
  \label{fig:cannon-alg}
\end{figure}

\subsection{Johnson's Algorithm}
Johnson's algorithm is a 3D algorithm, so it has a different target machine and initial data distribution, as seen in \autoref{fig:johnson-alg}.
Given $p$ processors, Johnson's algorithm targets a processor cube with side length
$\sqrt[3]{p}$: $\mM = \textsf{Grid}(\sqrt[3]{p},\sqrt[3]{p},\sqrt[3]{p})$.
The input matrices are partitioned into tiles, and then fixed to a face of the processor cube.
We express the distribution of $A$ using $\tdistname{A}{xy}{xy0}$,
which partitions $A$ by the first two dimensions of $\mM$, and then fixes the partition onto a face of
$\mM$.
The placements for $B$ and $C$ are similar, but restrict the matrices to different faces of the processor cube.

\begin{figure}%
      \begin{lstlisting}[mathescape=tru,frame=single,commentstyle=\color{gray},language=none,basicstyle=\ttfamily\scriptsize]
# Arrange $p$ processors into a 3D grid, $\sqrt[3]{p} \times \sqrt[3]{p} \times \sqrt[3]{p}$.
# Assign a tile of $A$ to each processor $P_{ij0}$.
# Assign a tile of $B$ to each processor $P_{i0k}$.
# Assign a tile of $C$ to each processor $P_{0jk}$.
for all $P_{ijk}$ in parallel:
  $P_{i0k}$ broadcasts $B_{ik}$ to each $P_{ijk}$
  $P_{0jk}$ broadcasts $C_{jk}$ to each $P_{ijk}$
  $A_{ijk}$ = $B_{ik}$ $\times$ $C_{kj}$
  $P_{ijk}$ sum reduces $A_{ijk}$ to $P_{ij0}$
      \end{lstlisting}
      \caption{Pseudocode for Johnson's algorithm.}
      \label{fig:johnson-alg}
\end{figure}

The schedule for Johnson's algorithm first distributes all dimensions of the iteration space with
\lstinline[mathescape=tru]!distribute({i,j,k}, {io,jo,ko}, {ii,ji,ki}, Grid($\sqrt[3]{p}$,$\sqrt[3]{p}$,$\sqrt[3]{p})$)! and
communicates under the distributed loop with \lstinline!communicate({A,B,C}, kn)!.
Every processor performs a local multiplication using corresponding chunks of $B$ and $C$, and reduces into
$A$, matching the original presentation of Johnson's algorithm.

\subsection{PUMMA, Solomonik's Algorithm, and COSMA}
The remaining algorithms in \autoref{fig:algorithm-schedules} can be derived using the
same principles.
The PUMMA algorithm is a hybrid between Cannon's algorithm and the SUMMA algorithm, using a broadcast to
communicate one of the matrices and a systolic pattern to communicate the other.
Solomonik's algorithm operates on a processor cube, where each slice performs
Cannon's algorithm on pieces of $B$ and $C$, and reduces the results of each slice into tiles of $A$.
The schedule for Solomonik's algorithm is very similar to the 2D algorithms---it
also \lstinline!distribute!-s over the $k$ dimension, and \lstinline!divide!-s the resulting
inner $ki$ loop into chunks.
COSMA derives a schedule that includes a machine organization, and a strategy for distributing
the $i$, $j$, and $k$ loops of the computation.
Given these parameters, our technique can generate code that implements the distribution layer of COSMA.%
\footnote{The COSMA algorithm additionally is able to split an iteration space dimension sequentially.
To express sequential splits, the outer dimensions must be \lstinline!divide!-ed,
and then inner parallel splits can be \lstinline!distribute!-ed.}

\section{Compilation}
\label{sec:ir}
To implement the distributed scheduling commands in an intermediate representation (IR) that can be reasoned
about by a compiler, we use the \emph{concrete index notation} IR
developed by Kjolstad et al.~\cite{taco_workspaces} and Senanayake et al.~\cite{taco_scheduling}.

\begin{figure}
\vspace*{-1em}
\footnotesize
\[
\begin{array}{rlrlrl}
\textit{Index Variables} & i & \textit{Constants} & c & \textit{Tensors} & \mT\\
\end{array}
\]
\vspace*{-1.5em}
\[
\begin{array}{rlcl}
  \textit{Accesses} & a & \bnfdef & \mT(i*) \\
  \textit{Expressions} & e & \bnfdef & a \bnfalt c \bnfalt e + e \bnfalt \ldots \\
  \textit{Scheduling Relation} & r & \bnfdef & \textsf{divide}(i, i_o, i_i, c) \bnfalt \ldots \\
  \textit{Statements} & S & \bnfdef & \forall_i~S \bnfalt a = e \bnfalt a \pluseq e \bnfalt \\
  & & & S~;~S \bnfalt S~\textsf{s.t.}~r* \\
\end{array}
\]
\caption{Syntax for Concrete Index Notation}
\label{fig:concrete-index-notation-syntax}
\end{figure}

\subsection{Concrete Index Notation}

Concrete index notation is a lower-level IR than tensor index notation that specifies the ordering
of \lstinline!for! loops, and tracks applied optimizations and loop transformations.
Tensor index notation statements are lowered into concrete index notation by constructing a loop nest
based on a left-to-right traversal of the variables in the tensor index notation statement.
The syntax for concrete index notation is shown in \autoref{fig:concrete-index-notation-syntax}.
Concrete index notation is further lowered into an imperative IR by target specific backends.

Scheduling transformations are expressed through rewrite rules on concrete index notation by
rewriting loops and tracking transformations through the \textsf{s.t.} clause.
An example transformation rule for the \lstinline!divide! command is
\[
  \ldots~\forall_i~S \xrightarrow{\textsf{divide}(i, i_o, i_i, c)} \ldots~\forall_{i_o}\forall_{i_i}~S~\textsf{s.t.}~\textsf{divide}(i, i_o, i_i, c)
\]
The full set of scheduling operations supported in TACO is described by Senanayake et al.~\cite{taco_scheduling}.

\subsection{Distributed Scheduling}

We now describe how each of the distributed scheduling commands transform concrete index notation statements.

\textbf{\lstinline!distribute!.} The \lstinline!distribute! transformation marks a loop
as distributed
for a backend specific pass to elaborate further.
\[
  \ldots~\forall_i~S \xrightarrow{\textsf{distribute}(i)} \ldots~\forall_i~S~\textsf{s.t.}~\textsf{distribute}(i)
\]

\textbf{\lstinline!rotate!.} Similarly to \lstinline!distribute!, \lstinline!rotate! is also expressed
as a transformation that adds the \textsf{rotate} relation to a statement.
\[
  \ldots~\forall_{I}\forall_t~S \xrightarrow{\textsf{rotate}(t, I, r)} \ldots\forall_{I}\forall_{r}~S~\textsf{s.t.}~\textsf{rotate}(t, I, r)
\]
\sloppy Rotation is implemented by setting $t = r + \sum I~\textsf{mod extent}(t)$,
offsetting the starting point of $t$.

\textbf{\lstinline!communicate!.} The \lstinline!communicate! command aggregates the communication
of data necessary for a set of loop iterations to the executing processor.
The \textsf{s.t.} clause of the $\forall$ targeted by a \lstinline!communicate! statement
stores all tensors which must be communicated at the $\forall$.
\[
  \ldots~\forall_i~S \xrightarrow{\textsf{communicate}(\mT, i)} \ldots~\forall_i~S~\textsf{s.t.}~\textsf{communicate}(\mT, i)
\]

The target backend controls how to further lower a $\forall$ with
\textsf{communicate} relations, by introducing logic to communicate accessed
components for each iteration of the $\forall$.

\subsection{Lowering Tensor Distribution Notation}
We implement the placement of a tensor into the distribution described by a 
tensor distribution notation statement by translating the into a concrete index 
notation statement that accesses data in the described orientation.
The translation algorithm is mechanical and uses the following steps for a distribution
$\tdist{X}{Y}$, where $X$ and $Y$ are sets of dimension names.
The algorithm is extended to hierarchical tensor distributions by applying the same idea
for each distribution level.
\begin{enumerate}
  \item Let $V$ be a set of index variables for each name in $X \cup Y$.
  \item Construct a concrete index notation statement $S$ of nested $\forall$
        loops for each variable in $V$. At the innermost loop, $S$ accesses
        $\mT$ with variables in $V$ corresponding to names in $X$. $\forall$'s corresponding
        to dimensions in $\mM$ fixed to a value are restricted to that value.
  \item \lstinline!reorder! $\forall$'s in $S$ such that the variables for $Y$ are the shallowest
        in the loop nest.
  \item \lstinline!divide! each index variable for $X$ by the corresponding dimension of $\mM$, and
        \lstinline!distribute! the outer variable.
  \item \lstinline!communicate! $\mT$ underneath all distributed variables.
\end{enumerate}
Intuitively, this procedure has two steps: 1) construct an iteration space over $\mT$ and any
broadcasted dimensions of $\mM$ and 2) distribute the iteration space onto $\mM$.
For example, the concrete index notation statement for $\tdist{xy}{x}$ is $\forall_{xo}\forall_{xi}\forall_{y}~\mT(x, y)~\textsf{s.t.}~\textsf{divide(x, xo, xi, gx)},~\allowbreak\textsf{distribute(xo)},~\textsf{communicate(}\mT\textsf{, xo)}$.

\section{Implementation}
\label{sec:implementation}
We target the Legion\cite{legion} distributed task-based runtime system.
Legion implements several features necessary for high performance on modern machines,
but orthogonal to the topics we discuss, including 1) overlap of communication and computation,
2) data movement through deep memory hierarchies, 3) native support for accelerators and 4)
control over placement of data and computation in target memories
and processors.
Therefore, our strategies for further lowering of concrete index notation are directed
by Legion's API.
Legion performs dynamic analysis to facilitate communication between disjoint memory spaces.
We discuss strategies for static communication analysis that are
compatible with our approach in \autoref{sec:related-work}.
We focus in this section on lowering concepts related to distribution and communication.
For lowering of sub-statements that execute on a single CPU or GPU, we follow the same process used in TACO.

\subsection{Legion Programming Model}

\emph{Regions} are Legion's abstraction for distributed data structures.
Regions can be viewed as multi-dimensional arrays, and we use them to represent dense tensors.

\emph{Tasks} are the unit of computation in Legion.
Tasks operate on regions, and the runtime system is responsible for moving data that a task requires into
a memory accessible by the processor the task is running on before the task begins execution.
When launching a task, users provide information to the runtime system naming the regions on which the task operates.
Multiple independent instances of the same task can be launched in a single operation called an {\em index task launch},
which is similar to a \texttt{parallel for} construct.

Regions can be \emph{partitioned} into subregions that can be operated on in parallel by tasks.
Legion has several ways to create partitions, including an API that uses hyper-rectangular
bounding boxes to partition regions into subregions.

Legion's \emph{mapping} interface allows for control over aspects of execution
such as which processors tasks execute on and which memories regions are allocated
in.

Communication in Legion is implicit.
The data desired to be communicated is described to Legion through application created partitions.
Legion then handles the physical movement of this data through channels specialized for the source and
destination memories, such as NVLink for GPU-GPU communication and GASNet-EX for inter-node communication.

\subsection{Lowering to Legion}

Our lowering process to Legion is guided by scheduling relations in the concrete index notation.
$\forall$'s tagged as distributed are lowered into index task launches over the
    extent of the loop, where the loop bodies are placed within Legion tasks.
    Directly nested distributed loops are flattened into multi-dimensional index
    task launches.
Legion partitions are created for each tensor denoted to \lstinline!communicate! under a loop.
    The bounds of the hyper-rectangles to use in the partitioning API are derived using a standard bounds analysis
    procedure using the extents of index variables.
Directives about processor kinds and machine grids are communicated to a mapper that places tasks in the desired processor orientations.

\section{Evaluation}
\label{sec:evaluation}

\textbf{Experimental Setup.} We ran our experiments on the Lassen~\cite{lassen} supercomputer.
Each Lassen node has a dual socket IBM Power9 CPU with 40 available cores.
Each node contains four NVIDIA Volta V100 GPUs connected by NVLink 2.0 and an
Infiniband EDR interconnect.
All code was compiled with GCC 8.3.1 \verb!-O3! and CUDA 11.1.
Legion\footnote{\url{https://gitlab.com/StanfordLegion/legion/}, commit 8ca3331c5d.} was configured with
GASNet-EX 2021.3.0 for communication.%

\textbf{Comparison Targets.} We compare against ScaLAPACK as provided by LAPACK on the Lassen system
(version 3.9.0), Cyclops Tensor Framework\footnote{\url{https://github.com/cyclops-community/ctf}, commit 36b1f6de53},
and the original COSMA\footnote{\url{https://github.com/eth-cscs/COSMA/}, commit c7bdab95ba} implementation.
All systems were configured to use OpenBLAS\footnote{\url{https://github.com/xianyi/OpenBLAS/}, commit 37ea8702e} compiled
with OpenMP, and (if supported) the CuBLAS shipped with CUDA 11.1 for local BLAS operations.

\textbf{Overview.}
To evaluate the absolute performance of our system, we present comparisons
of CPU and GPU matrix-multiplication performance between \name{} and existing libraries.
We then consider several higher order tensor computations to show that our system
achieves good performance on kernels that receive less attention from researchers.
These comparisons collectively demonstrate that \name{} achieves high 
absolute performance, comparable to hand-tuned codes, and that \name{}'s 
abstractions generalize to optimization of higher order tensor kernels.

\subsection{Distributed Matrix-Multiplication Benchmarks}

\begin{figure*}
    \begin{subfigure}{0.5\textwidth}
        \centering
        \includegraphics[width=\textwidth]{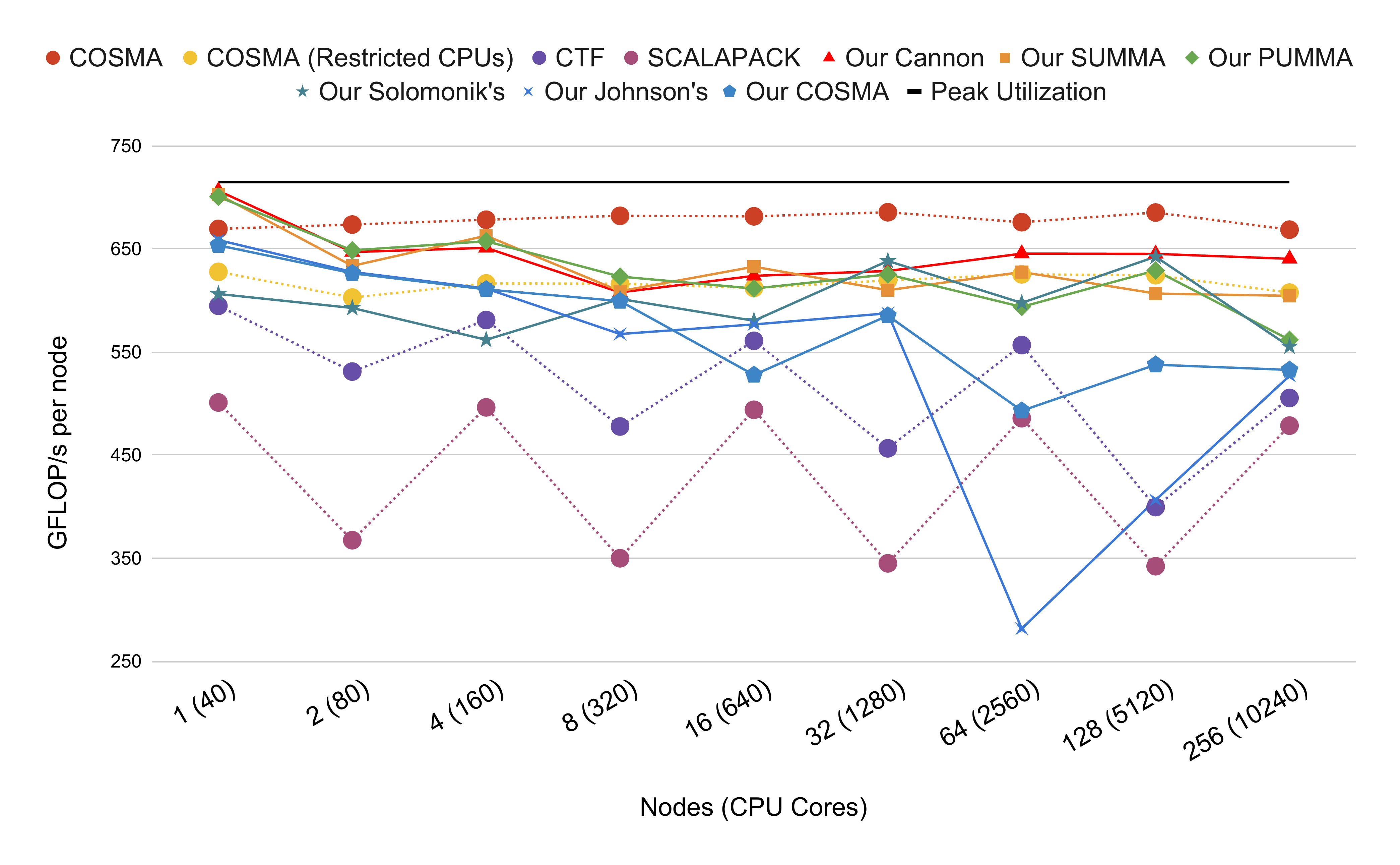}
        \caption{CPU results.}
        \label{fig:cpu-weak-scaling}
    \end{subfigure}%
    \begin{subfigure}{0.5\textwidth}
        \centering
        \includegraphics[width=\textwidth]{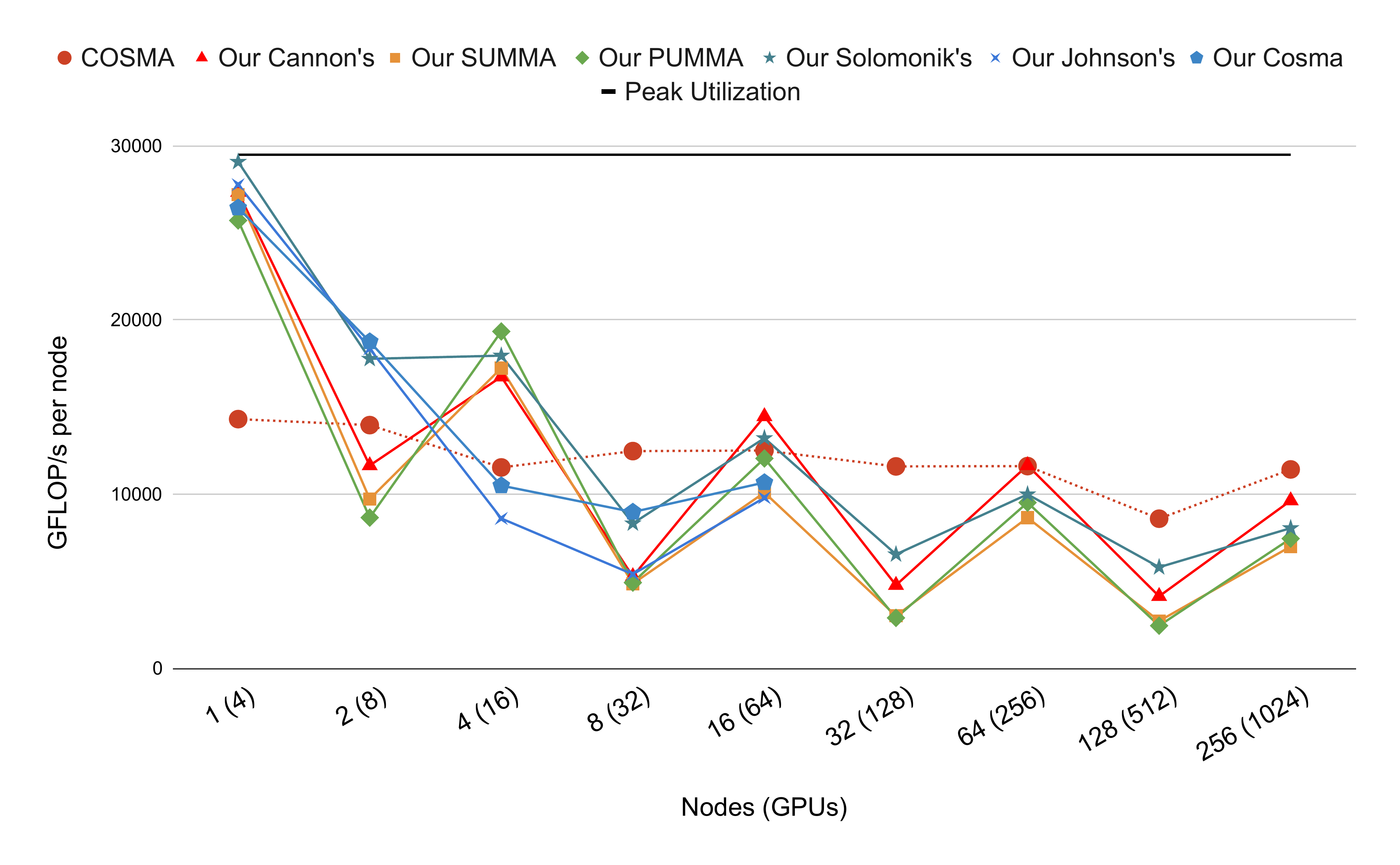}
        \caption{GPU results.}
        \label{fig:gpu-weak-scaling}
    \end{subfigure}
    \caption{Weak-scaling results (higher is better) for matrix-multiplication. \name{}'s kernels are prefixed with "Our" and are denoted with filled lines. Other systems (COSMA, CTF and ScaLAPACK) are denoted with dotted lines.}
    \label{fig:gemm-results}
\end{figure*}

We evaluate \name{}'s implementations of distributed matrix-multiplication by
comparing against ScaLAPACK~\cite{scalapack}, Cyclops Tensor Framework (CTF)~\cite{ctf}
and COSMA~\cite{cosma}.
ScaLAPACK is a well-known library that provides a variety of distributed kernels
for many common linear algebra operations.
CTF is a library for distributed tensor algebra that implements the 2.5D matrix
multiplication algorithm of Solomonik et al.~\cite{solomonikMM}.
Finally, COSMA is recent work that provides both a theoretically optimal algorithm
as well as the currently best known dense matrix-multiplication implementation.
Of these systems, only COSMA has a GPU backend, so we restrict our comparison
to COSMA in the GPU setting.\footnote{CTF advertises GPU support, but we were not able to successfully build it.}

For comparison, we implement all algorithms discussed in \autoref{sec:algorithms}, including
the SUMMA algorithm used by ScaLAPACK, the 2.5D algorithm used by CTF, and the COSMA algorithm.
Results for CPUs and GPUs are in \autoref{fig:gemm-results}.
These experiments are weak-scaled (memory per node stays constant) on square matrices,
where the initial problem sizes were 8192x8192 and 20000x20000 for CPUs and GPUs respectively.
Initial problem sizes were chosen to be just large enough to achieve peak utilization on a single node.

For CPUs, we find COSMA performs best with 40 ranks per node, while ScaLAPACK and CTF
perform best with 4 ranks per node.
For GPUs, we run COSMA with one rank per GPU.
\name{}'s kernels are run with one rank per node.

\subsubsection{CPU Results}

\autoref{fig:cpu-weak-scaling} contains the CPU distributed matrix-multiplication benchmarks.
At 256 nodes, CTF and ScaLAPACK achieve at most 80\% performance of
codes generated by our system and COSMA. They also experience performance variability
due to effects from non-square machine grids.
COSMA and \name{} achieve higher performance by overlapping communication and computation more effectively---our profiles show that for CPUs, it is possible to hide nearly all communication costs with computation.
Since most communication costs can be hidden, we do not find significant performance variations between the
different algorithms implemented in \name{}, except for Johnson's algorithm, which experiences performance
degradation on processor grids that aren't perfect cubes (we model each CPU socket as an abstract \name{} processor).
Finally, we see that our best schedules are within 10\% of the performance of COSMA on all node counts,
and within 5\% on 256 nodes.
\name{} has a penalty compared to COSMA because we allocate 4 CPU cores per node to Legion to
perform runtime dependence analysis and utility work.\footnote{The number of runtime cores is a configurable parameter. We found the best performance with 4 cores per node.}
The line named "COSMA (Restricted CPUs)" in \autoref{fig:cpu-weak-scaling} shows that COSMA achieves
equal performance to \name{} when restricted to use 36 out of the 40 CPU cores, 
which is number of work cores allocated to \name{}.
Although use of the Legion runtime imposes a small cost (5-10\%), we believe that this cost is worth
paying for simpler engineering and improved programmer productivity.

\subsubsection{GPU Results}\label{sec:gpu-gemm}

\autoref{fig:gpu-weak-scaling} contains the GPU distributed matrix-multiplication benchmarks.
On a single node, all of our kernels achieve twice the performance of COSMA, and COSMA achieves 15\% higher
performance on 256 nodes than \name{}'s best performing schedule.
COSMA keeps all data in CPU memory and uses an out-of-core GEMM kernel
that pulls data into the GPU for computation with CuBLAS.
\name{}'s kernels keep all data in GPU framebuffer memory and communicate via NVLink,
achieving near-peak utilization on a single node\footnote{The COSMA developers did not have access to machines with NVLink during development and thus do not include NVLink support~\cite{gregory_comm}. NVLink support for COSMA is an engineering limitation and not fundamental.}.

In contrast to the CPU experiments, we see different performance characteristics between algorithms
when moving to multiple nodes.
The larger problem sizes required to achieve peak GFLOP/s on a single GPU cause 
the computation to be evenly balanced between communication and computation, 
and therefore extremely sensitive to the communication costs of each algorithm.

The 2D algorithms (Cannon, SUMMA and PUMMA) perform well at square node counts (equaling or exceeding the performance of COSMA),
and perform worse at rectangular node counts, similar to the variations seen with ScaLAPACK and CTF on CPUs.
This variation comes from the imbalanced communication in the rectangular case.
Within the 2D algorithm family, we see that Cannon's algorithm outperforms SUMMA and PUMMA as
the node count increases.
The difference is the systolic communication pattern enabled by \lstinline!rotate! that Cannon's
algorithm uses.
Avoiding the collective-style broadcasting operations and using nearest-neighbor communication only, our schedule
using Cannon's algorithm achieves higher performance at scale.

The 3D algorithms (Johnson's, Solomonik's 2.5D and COSMA) trade communication for
extra memory use, and achieve higher performance on the non-square node counts than their 2D counterparts.
Johnson's algorithm achieves high performance when the number of GPUs is a perfect cube, but for non-cubes achieves worse
performance due to over-decomposition.
Our implementation of COSMA achieves better performance than Johnson's algorithm because it can adapt the
decomposition to the machine size, but does not equal the performance of the 2D algorithms due to the lack of matrix-multiplication
specialized broadcast and reduction operators as in the COSMA author's implementation.
Johnson's algorithm and our COSMA implementation run out of memory at 32 nodes, because they replicate input components onto
multiple nodes, they exhaust the limited GPU memory at higher node counts. The COSMA author's implementation use the larger CPU memory to hold matrices.
Solomonik's 2.5D algorithm interpolates between the 2D and 3D algorithms, using extra memory when possible to
speed up the computation.
In our implementation, we utilize extra memory on the non-square node counts, resulting in better performance
than the 2D algorithms on those processor counts.

\name{}'s kernels perform worse than COSMA and experience larger performance variations due to communication costs.
Legion's DMA system is unable to achieve peak bandwidth out of a node (18/25 GB/s) when data is resident
in the GPU framebuffer memory, while the system (and COSMA) achieve near peak when data is resident in CPU memory.
The Legion team plans to address this shortcoming in the future.

\subsection{Higher Order Tensor Benchmarks}

\begin{figure*}
    \begin{subfigure}{0.5\textwidth}
        \centering
        \begin{subfigure}{0.5\textwidth}
            \centering
            \includegraphics[width=\textwidth]{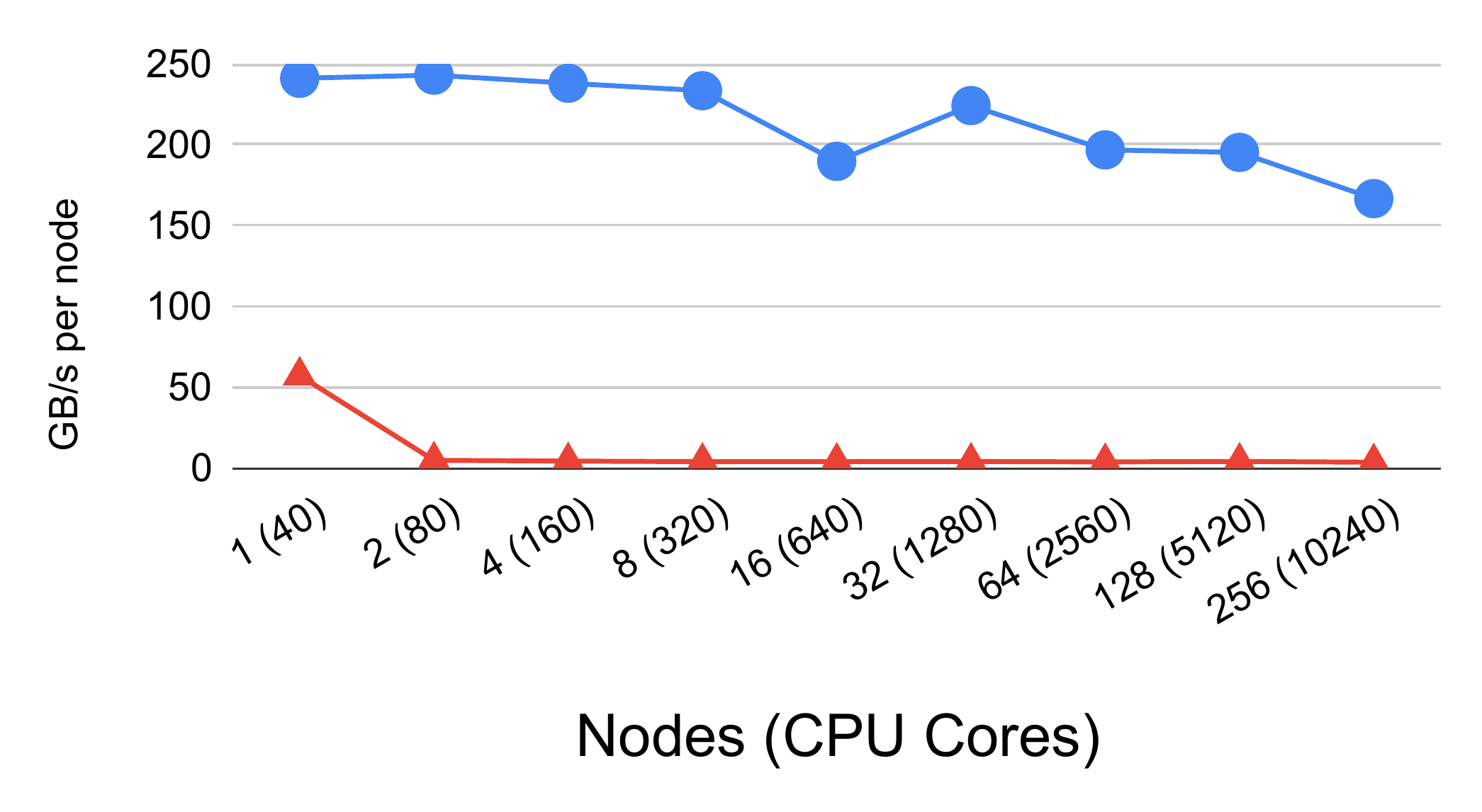}
        \end{subfigure}\hfill
        \begin{subfigure}{0.5\textwidth}
            \centering
            \includegraphics[width=\textwidth]{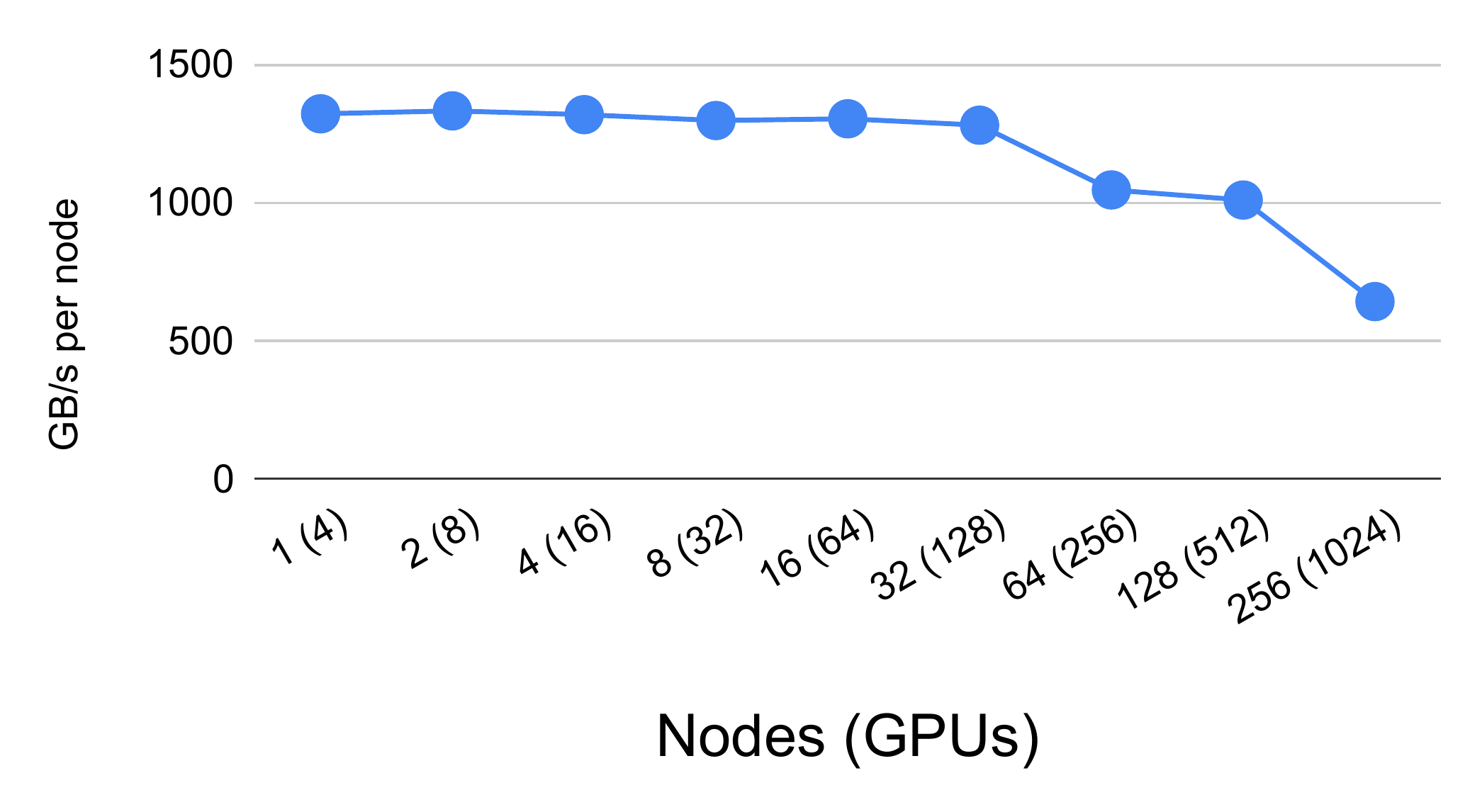}
        \end{subfigure}
        \caption{TTV}
        \label{fig:ttv}
    \end{subfigure}\hfill
    \begin{subfigure}{0.5\textwidth}
        \centering
        \begin{subfigure}{0.5\textwidth}
            \centering
            \includegraphics[width=\textwidth]{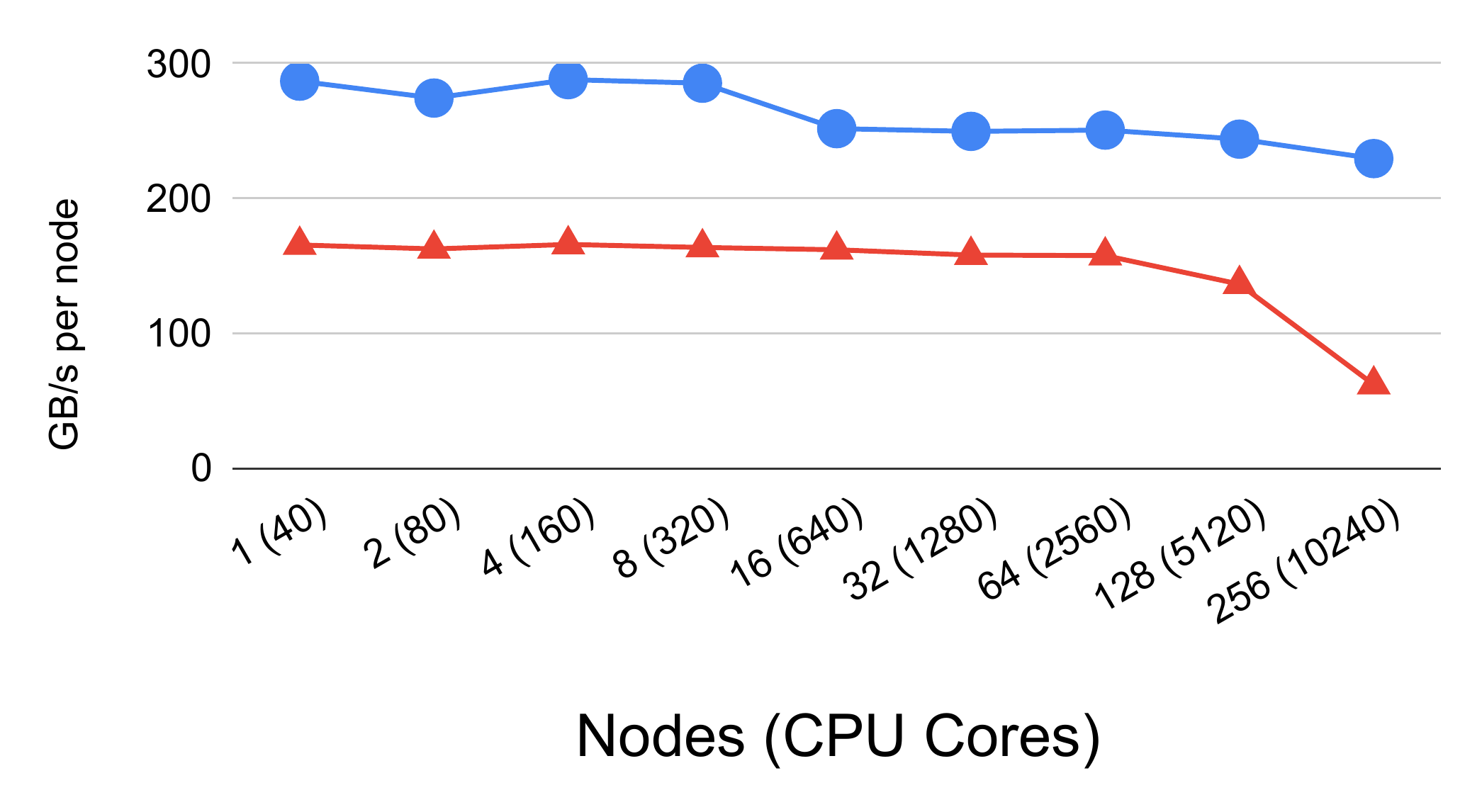}
        \end{subfigure}\hfill
        \begin{subfigure}{0.5\textwidth}
            \centering
            \includegraphics[width=\textwidth]{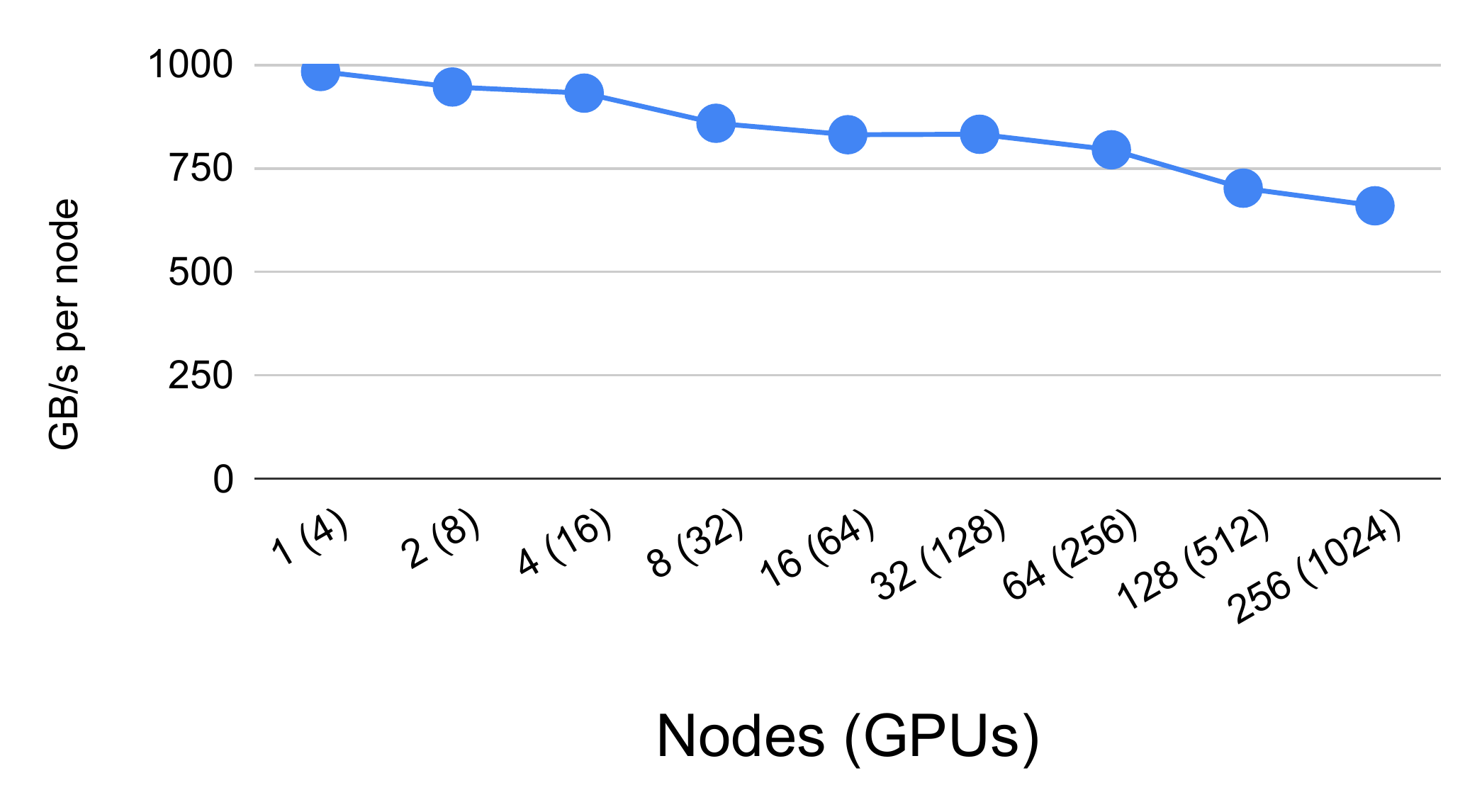}
        \end{subfigure}
        \caption{Innerprod}
        \label{fig:innerprod}
    \end{subfigure}

    \begin{subfigure}{0.5\textwidth}
        \centering
        \begin{subfigure}{0.5\textwidth}
            \centering
            \includegraphics[width=\textwidth]{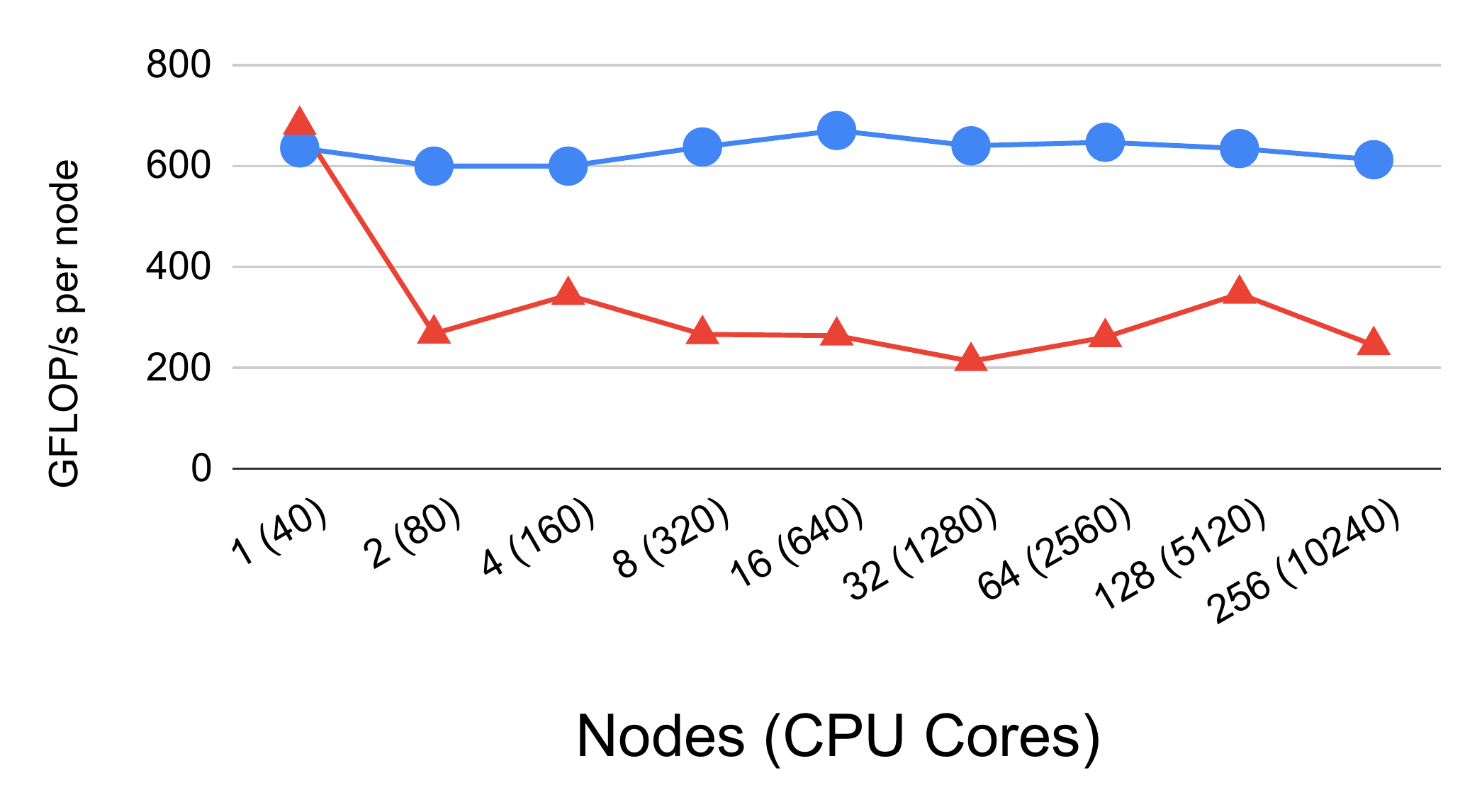}
        \end{subfigure}\hfill
        \begin{subfigure}{0.5\textwidth}
            \centering
            \includegraphics[width=\textwidth]{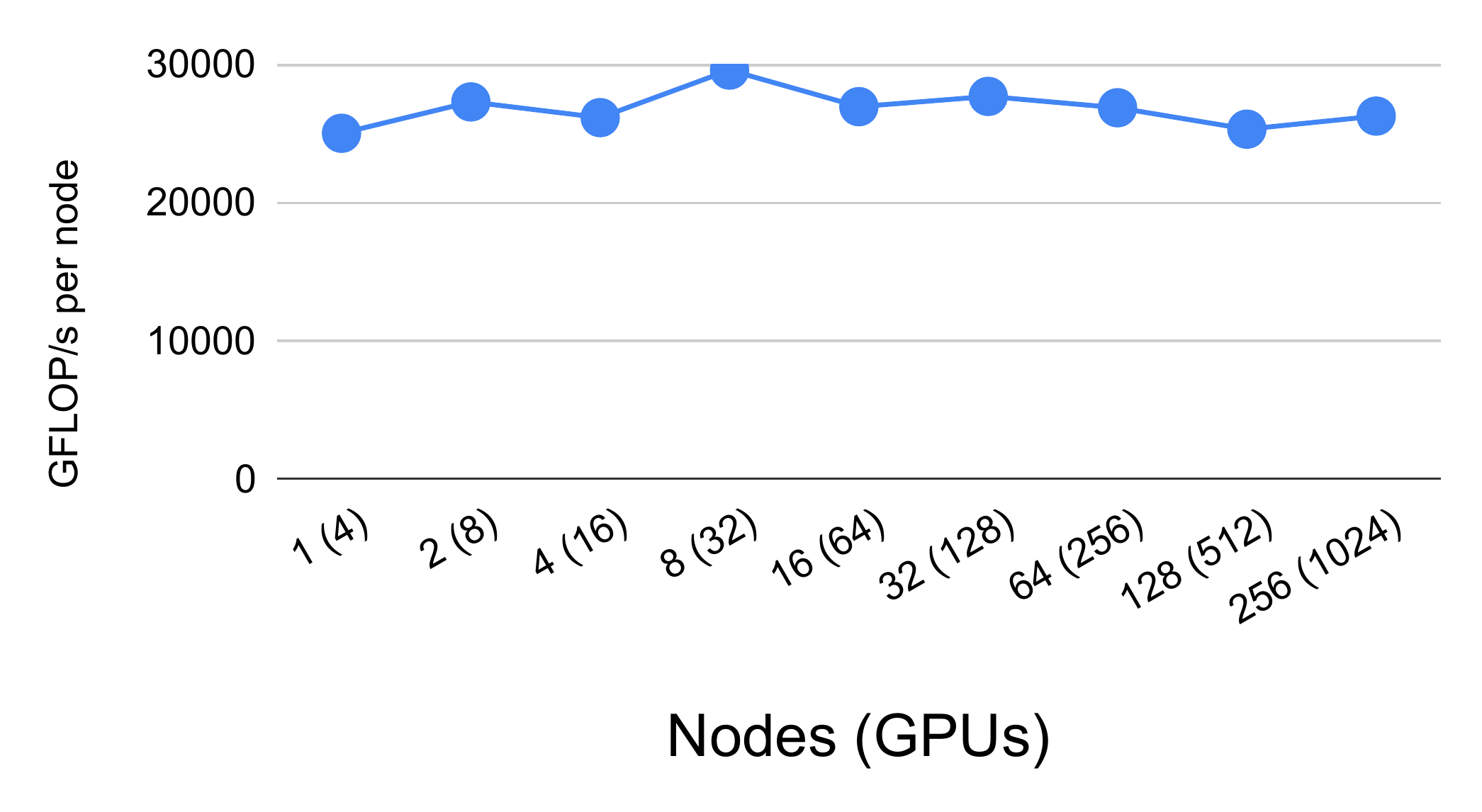}
        \end{subfigure}
        \caption{TTM}
        \label{fig:ttm}
    \end{subfigure}\hfill
    \begin{subfigure}{0.5\textwidth}
        \centering
        \begin{subfigure}{0.5\textwidth}
            \centering
            \includegraphics[width=\textwidth]{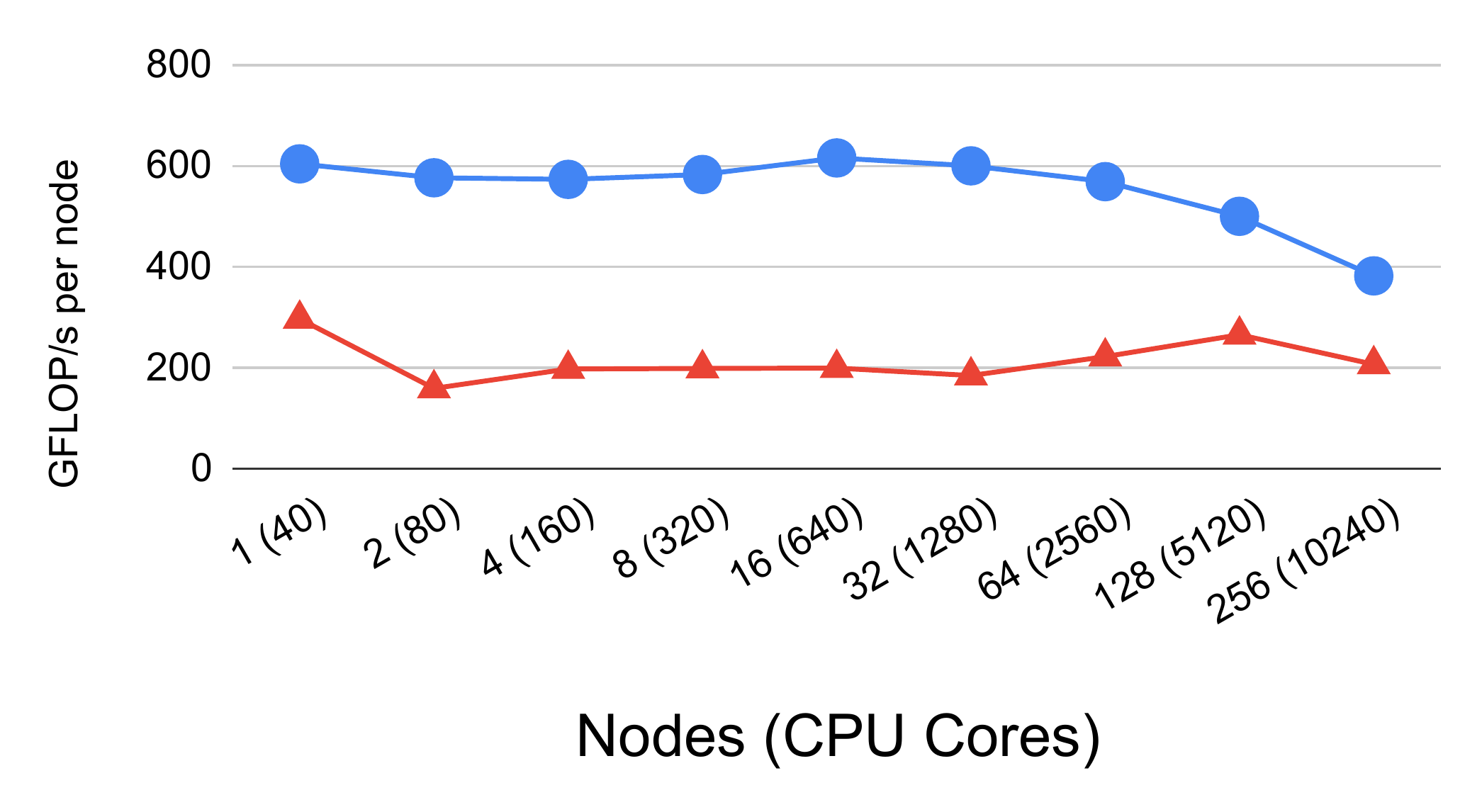}
        \end{subfigure}\hfill
        \begin{subfigure}{0.5\textwidth}
            \centering
            \includegraphics[width=\textwidth]{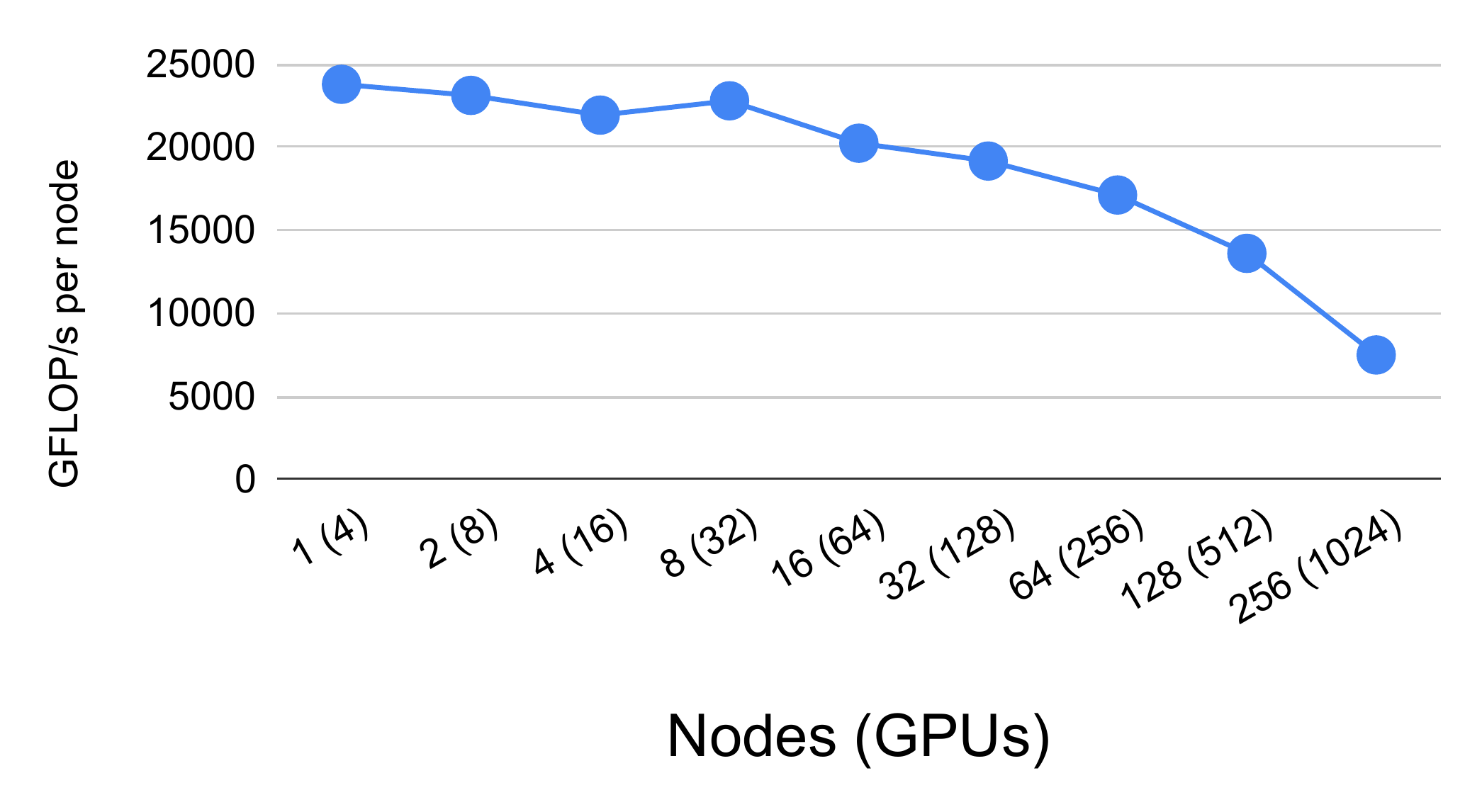}
        \end{subfigure}
        \caption{MTTKRP}
        \label{fig:mttkrp}
    \end{subfigure}

    \begin{subfigure}{\textwidth}
        \centering
        \vspace*{-1em}
        \includegraphics[width=0.10\textwidth]{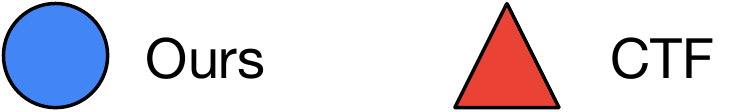}
    \end{subfigure}
    \caption{Weak-scaling results (higher is better) of higher order tensor computations.}%
    \label{fig:higher-order}
\end{figure*}

To evaluate the generality of our system, we compare against the Cyclops Tensor Framework (CTF), which is the only
system that we know of that offers similar generality: distributed implementations of
any tensor algebra operation.
We consider the following tensor expressions:%
\begin{itemize}
    \item Tensor times vector (TTV): {\small $A(i, j) = B(i, j, k) \cdot c(k)$}
    \item Inner product (Innerprod): {\small $a = B(i, j, k) \cdot C(i, j, k)$}
    \item Tensor times matrix (TTM): {\small $A(i, j, l) = B(i, j, k) \cdot C(k, l)$}
    \item Matricized tensor times Khatri-Rao Product (MTT\-KRP): {\small $A(i, l) = B(i, j, k) \cdot C(j, l) \cdot D(k, l)$}
\end{itemize}
These operations all have real-world applications.
For example, the TTM and MTTKRP kernels are important building blocks in routines that compute
Tucker and canonical polyadic decompositions of tensors~\cite{baderkolda}.

CTF decomposes arbitrary tensor operations into calls to a distributed matrix-multiplication implementation
by slicing and reshaping the tensors.
While such a strategy can implement all tensor algebra operations, it cannot implement an optimal strategy
for every operation.
Our approach of generating a bespoke implementation for a target kernel allows for development of
a schedule that implements an optimal strategy for each kernel.
Our experiments show that we outperform CTF on each higher-order tensor expression that we consider.
The tradeoff between these systems is that CTF fully automates the distribution process, while users
must provide a schedule to distribute their computations in \name{}.
We note that the scheduling primitives in \name{} provide a mechanism for
future work to target when automatically schedule computations for distribution.

Weak-scaling (memory per node stays constant) experiments for CPUs and GPUs are shown in \autoref{fig:higher-order}.
For kernels that are bandwidth bound (TTV and innerprod), we report results in GB/s, rather than GFLOP/s.
We run CTF with 4 ranks per node, for all kernels other than innerprod, where we found best performance at 40 ranks per node.
Because we were unable to build CTF's GPU backend, we only report CPU results.
For each higher order kernel, we either experimented with different schedules to minimize
inter-node communication (TTV, innerprod, TTM) or implemented a known algorithm (MTTKRP~\cite{ballardmttkrp}).
Input tensors were laid out in a row-major layout and distributed in a manner that matched the chosen schedule
or as specified by a proposed algorithm (MTTKRP).
We used the same data distributions and distribution schedules for the CPU and GPU kernels.
As with the matrix-multiplication benchmarks, initial problem sizes were chosen to be just large enough
to achieve peak utilization on a single node.

\subsubsection{Scheduling Leaf Kernels}
Leaf kernel performance heavily impacts the overall performance of a distributed
computation.
To keep the leaf kernels similar between \name{} and CTF, we use the same
strategy as CTF on a single node and cast the TTM and MTTKRP operations to loops of matrix-multiplications.
For element-wise operations (TTV and innerprod), we parallelize and vectorize
loops for CPUs, and tile loops to thread blocks for GPUs.
CTF aims at scalability to large core counts rather than fully utilizing the resources
on a single node.
This approach results in worse single-node performance for the TTV, innerprod and MTTKRP benchmarks (when casting
MTTKRP to matrix-multiplications, an element-wise reduction operation is required).

\subsubsection{Results}

\textbf{TTV (\autoref{fig:ttv}).}
Casting the TTV operation as a sequence of matrix-multiplication operations performs unnecessary communication
and CTF's performance drops past a single node.
Instead, our schedule using \name{} performs the operation element-wise without communication.
Our GPU kernel achieves higher bandwidth than the CPU kernel, but starts to fall off
at 256 GPUs due to the kernels' short execution time (several milliseconds).

\textbf{Innerprod (\autoref{fig:innerprod}).}
The inner-product kernel is best implemented as a node-level reduction followed by a global reduction over all nodes,
and we use this strategy in the \name{} schedule.
CTF achieves good weak scaling performance as well, but is still slower than our implementation using \name{}.
The GPU implementation of innerprod has similar performance characteristics as TTV.

\textbf{TTM (\autoref{fig:ttm}).}
Our DISTAL schedule for TTM expresses the kernel as a set of parallel matrix-multiplication operations by
\lstinline!distribute!-ing the $i$ loop of the kernel.
This strategy results in no inter-node communication and both our
CPU and GPU implementations achieve high efficiency up to 256 nodes.
Instead, CTF casts the kernel as a loop of distributed matrix-multiplications, which results in a large
drop in performance due to inter-node communication.

\textbf{MTTKRP (\autoref{fig:mttkrp}).}
\name{} allows for implementing specialized algorithms for tensor algebra operations.
We implement the algorithm of Ballard et al.~\cite{ballardmttkrp} that keeps the
3-tensor in place and reduces intermediate results into the output tensor.
\name{}'s kernels fall off after 64 nodes due to overheads of algorithms used within
Legion to manage the situation where portions of regions are replicated onto many nodes.
The Legion team plans to address this shortcoming in the future.
CTF does not achieve similar performance on a single node, but has flat scaling behavior.

\section{Related Work}
\label{sec:related-work}
\textbf{Distributed Tensor Algebra.}
The only system that we know of to support distributed execution of any tensor algebra
kernel is the Cyclops Tensor Framework (CTF)~\cite{ctf}.
CTF casts tensor contractions into a series of distributed matrix-multiplication
operations and transposes, an approach also used by the single-node Tensor Contraction
Engine~\cite{tensorcontractionengine}.
Our approach of generating specialized kernel implementations
can lead to improvements over CTF's interpreted approach.

Distributed algorithms for tensor algebra have drawn interest from researchers,
including the matrix-multiplication algorithms
in Figure~\ref{fig:algorithm-schedules} and algorithms for higher order tensor algebra like
MTTKRP~\cite{ballardmttkrp}.
\name{} provides a framework to model and generate implementations for these algorithms.

Related work in the database community has taken steps to extend relational database
engines to support distributed linear algebra~\cite{jermaine-dist-lin-alg} and
distributed tensor computations~\cite{jermaine-dist-tensors}.

\textbf{DSL Compilers.}
Several DSL compilers for single-node systems have been developed, such as Halide~\cite{halide},
TVM~\cite{TVM}, Tensor Comprehensions~\cite{tensorcomprehensions}, COGENT~\cite{cogent} and TACO~\cite{taco}.
Distributed Halide~\cite{dist-halide} and Tiramisu~\cite{tiramisu} support targeting distributed machines.
Distributed Halide extends Halide with data and computation distribution commands.
\name{} supports richer data distributions (such as broadcasting and fixing tensor partitions)
and targets tensor algebra instead of stencil codes. %
Tiramisu is a polyhedral compiler that can target distributed machines
and supports scheduling commands that distribute computation and communicate data.
However, these commands require more user input: users must describe the
data to communicate and the processors involved.
Recent work by Ihadadene~\cite{tiramisu-comm} automates these components for stencil codes.
Unlike \name{}, Tiramisu does not have a data distribution language or a \lstinline!rotate! command, which limits 
its ability to generate some sophisticated distributed algorithms.
\name{} is the first system of this kind to separate data and computation distributions, allowing for
expression of many existing distributed tensor algebra algorithms, including all of the algorithms
in \autoref{fig:algorithm-schedules}.

\textbf{Modeling Machines and Distributing Data.}
Common data distributions of arrays (such as row, column, and tiled) were first included as directives in languages for distributed programming (e.g., High Performance Fortran~\cite{high-performance-fortran}).
ZPL introduced the idea of separately defining an abstract machine (e.g., as a grid of processors) and a function defining a partitioning and mapping of data onto that machine \cite{zpl}. 
This approach has been adopted by
Chapel \cite{chapel} and extended in Sequoia (to a hierarchical abstract machine \cite{Sequoia}) and Legion \cite{legion} (to allowing multiple partitions of the same data or to be used simultaneously).

Dryden et al.~\cite{snir-notation} use a similar notation as \name{} that describes how
each dimension of a tensor is distributed to describe algorithms for distributed convolutional
neural network training.
Dryden et al. were inspired by the FLAME~\cite{elementalMM,schatzthesis} project,
which used a set notation to describe tensor distributions and redistribution
of tensors into different distributed layouts.
\name{} combines data distribution descriptions with a separate computation
scheduling language to allow for expression of many distributed algorithms.

\textbf{Distributed Polyhedral Compilation.}
Amarasinghe et al.~\cite{saman-dist} and Bondhugula~\cite{bondugula} used
polyhedral analysis to derive communication information for computation distribution from
affine loop nests statically.
Our work starts from a higher level representation that allows for the expression of
different algorithms through scheduling, whereas such decisions would need to
be already present in the affine loops targeted by these works.
The analysis of Amarasinghe et al. is fully static on a set of virtual processors, but can
result in imprecise communication when mapping the virtual processors onto
the physical processors.
In contrast, Bondhugula generates a set of runtime calls that complement
the static analysis to determine precise communication partners.
The ideas presented by Bondhugula and Amarasinghe et al. could be used as
analysis passes for an MPI-based backend for \name{} and are thus orthogonal to our approach.

\textbf{Discussion.} The critical difference between \name{} and prior work is the
notion of specifying independently how computation and data are distributed
in two high level languages.
The separation of these two concepts allows for flexibility in expression of
different algorithms and adaptability when integrating with existing codes---code can shape to data so that data may stay at rest.
Additionally, our combined static-dynamic approach allows for expression of
complex communication patterns and data distributions statically, while discharging
lower level data movement operations to a runtime system.
This design decision allows us to avoid complicated and, in some cases, brittle analyses
used by fully static approaches.

\section{Future Work}

We see many interesting avenues of future work from \name{}.
One such avenue is \name{}'s potential applications in training and
evaluating distributed deep learning models, where \name{} can be used
to generate distributed kernels for stages in the model.
As \name{} allows for separation of both data and computation distribution,
these parameters could be included in search-based approaches to deep
learning model distribution.
Another avenue is auto-scheduling and auto-formatting frameworks for \name{}.
Currently, \name{} is a useful productivity tool allowing for application developers
to develop code at a high level while performance engineers can optimize the mapping
without changing application code.
With automatic schedule and format selection, application developers could independently
achieve high performance and allow performance engineers to optimize further when an
automatic schedule is insufficient.
A third avenue of future work that we are currently undertaking is to
is to extend \name{} with support for sparse tensors.
The envisioned system would enable users to create distributed implementations 
of any desired tensor computation with any set of tensor formats.

\section{Conclusion}
We have introduced \name{}, a compiler for dense tensor algebra that targets modern,
heterogeneous machines.
\name{} allows for independent specifications of desired computation, data distribution,
and computation distribution.
The combination of data distribution and computation distribution allows for expression
of widely known algorithms and optimization of higher order tensor kernels.
\name{} generates code competitive with hand-optimized implementations of matrix-multiplication
and outperforms existing systems on higher order tensor kernels by between 1.8x and 3.7x.

\section*{Acknowledgements}
We would like to thank our anonymous reviewers, and especially our shepherd, for their valuable
comments that helped us improve this manuscript.
We would like to thank Olivia Hsu, Charles Yuan, Axel Feldmann, Elliot Slaughter and David Lugato for their comments on early stages of this manuscript.
We would like to thank the Legion team, including Mike Bauer, Sean Treichler, Manolis Papadakis, and Wonchan Lee for their feedback and support during the development of DISTAL.
We would like to thank the COSMA and CTF authors for their assistance in setup and benchmarking of their software.
Rohan Yadav was supported by an NSF Graduate Research Fellowship.
This work was supported in part by the Advanced Simulation and Computing (ASC) program of the US Department of Energy’s National Nuclear Security
Administration (NNSA) via the PSAAP-III Center at Stanford, Grant No. DE-NA0002373 and the Exascale Computing Project (17-SC-20-SC), a collaborative effort of the U.S. Department of Energy Office of Science and the National Nuclear Security Administration; the U.S. Department of Energy, Office of Science under Award DE-SCOO21516.
This work was also supported by the Department of Energy Office of Science, Office of Advanced Scientific Computing Research under the guidance of Dr. Hal Finkel.

\bibliography{main}


\begin{thebibliography}{37}


\ifx \showCODEN    \undefined \def \showCODEN     #1{\unskip}     \fi
\ifx \showDOI      \undefined \def \showDOI       #1{#1}\fi
\ifx \showISBNx    \undefined \def \showISBNx     #1{\unskip}     \fi
\ifx \showISBNxiii \undefined \def \showISBNxiii  #1{\unskip}     \fi
\ifx \showISSN     \undefined \def \showISSN      #1{\unskip}     \fi
\ifx \showLCCN     \undefined \def \showLCCN      #1{\unskip}     \fi
\ifx \shownote     \undefined \def \shownote      #1{#1}          \fi
\ifx \showarticletitle \undefined \def \showarticletitle #1{#1}   \fi
\ifx \showURL      \undefined \def \showURL       {\relax}        \fi
\providecommand\bibfield[2]{#2}
\providecommand\bibinfo[2]{#2}
\providecommand\natexlab[1]{#1}
\providecommand\showeprint[2][]{arXiv:#2}

\bibitem[\protect\citeauthoryear{Agarwal, Balle, Gustavson, Joshi, and
  Palkar}{Agarwal et~al\mbox{.}}{1995}]%
        {johnson}
\bibfield{author}{\bibinfo{person}{R.~C. Agarwal}, \bibinfo{person}{S.~M.
  Balle}, \bibinfo{person}{F.~G. Gustavson}, \bibinfo{person}{M. Joshi}, {and}
  \bibinfo{person}{P. Palkar}.} \bibinfo{year}{1995}\natexlab{}.
\newblock \showarticletitle{A three-dimensional approach to parallel matrix
  multiplication}.
\newblock \bibinfo{journal}{\emph{IBM Journal of Research and Development}}
  \bibinfo{volume}{39}, \bibinfo{number}{5} (\bibinfo{year}{1995}),
  \bibinfo{pages}{575--582}.
\newblock
\urldef\tempurl%
\url{https://doi.org/10.1147/rd.395.0575}
\showDOI{\tempurl}


\bibitem[\protect\citeauthoryear{Amarasinghe and Lam}{Amarasinghe and
  Lam}{1993}]%
        {saman-dist}
\bibfield{author}{\bibinfo{person}{Saman Amarasinghe} {and}
  \bibinfo{person}{Monica Lam}.} \bibinfo{year}{1993}\natexlab{}.
\newblock \showarticletitle{Communication Optimization and Code Generation for
  Distributed Memory Machines}.
\newblock \bibinfo{journal}{\emph{Sigplan Notices - SIGPLAN}}
  \bibinfo{volume}{28}, \bibinfo{pages}{126--138}.
\newblock
\urldef\tempurl%
\url{https://doi.org/10.1145/173262.155102}
\showDOI{\tempurl}


\bibitem[\protect\citeauthoryear{Baghdadi, Ray, Romdhane, Sozzo, Akkas, Zhang,
  Suriana, Kamil, and Amarasinghe}{Baghdadi et~al\mbox{.}}{2018}]%
        {tiramisu}
\bibfield{author}{\bibinfo{person}{Riyadh Baghdadi}, \bibinfo{person}{Jessica
  Ray}, \bibinfo{person}{Malek~Ben Romdhane}, \bibinfo{person}{Emanuele~Del
  Sozzo}, \bibinfo{person}{Abdurrahman Akkas}, \bibinfo{person}{Yunming Zhang},
  \bibinfo{person}{Patricia Suriana}, \bibinfo{person}{Shoaib Kamil}, {and}
  \bibinfo{person}{Saman Amarasinghe}.} \bibinfo{year}{2018}\natexlab{}.
\newblock \bibinfo{title}{Tiramisu: A Polyhedral Compiler for Expressing Fast
  and Portable Code}.
\newblock
\newblock
\showeprint[arxiv]{1804.10694}~[cs.PL]


\bibitem[\protect\citeauthoryear{Ballard, Knight, and Rouse}{Ballard
  et~al\mbox{.}}{2018}]%
        {ballardmttkrp}
\bibfield{author}{\bibinfo{person}{Grey Ballard}, \bibinfo{person}{Nicholas
  Knight}, {and} \bibinfo{person}{Kathryn Rouse}.}
  \bibinfo{year}{2018}\natexlab{}.
\newblock \showarticletitle{Communication Lower Bounds for Matricized Tensor
  Times Khatri-Rao Product}. In \bibinfo{booktitle}{\emph{2018 IEEE
  International Parallel and Distributed Processing Symposium (IPDPS)}}.
  \bibinfo{pages}{557--567}.
\newblock
\urldef\tempurl%
\url{https://doi.org/10.1109/IPDPS.2018.00065}
\showDOI{\tempurl}


\bibitem[\protect\citeauthoryear{Bauer, Treichler, Slaughter, and Aiken}{Bauer
  et~al\mbox{.}}{2012}]%
        {legion}
\bibfield{author}{\bibinfo{person}{Michael Bauer}, \bibinfo{person}{Sean
  Treichler}, \bibinfo{person}{Elliott Slaughter}, {and} \bibinfo{person}{Alex
  Aiken}.} \bibinfo{year}{2012}\natexlab{}.
\newblock \showarticletitle{Legion: Expressing Locality and Independence with
  Logical Regions}. In \bibinfo{booktitle}{\emph{Proceedings of the
  International Conference on High Performance Computing, Networking, Storage
  and Analysis}} (Salt Lake City, Utah) \emph{(\bibinfo{series}{SC '12})}.
  \bibinfo{publisher}{IEEE Computer Society Press},
  \bibinfo{address}{Washington, DC, USA}, Article \bibinfo{articleno}{66},
  \bibinfo{numpages}{11}~pages.
\newblock
\showISBNx{9781467308045}


\bibitem[\protect\citeauthoryear{Baumgartner, Auer, Bernholdt, Bibireata,
  Choppella, Cociorva, Gao, Harrison, Hirata, Krishnamoorthy, Krishnan, chung
  Lam, Lu, Nooijen, Pitzer, Ramanujam, Sadayappan, and Sibiryakov}{Baumgartner
  et~al\mbox{.}}{2005}]%
        {tensorcontractionengine}
\bibfield{author}{\bibinfo{person}{G. Baumgartner}, \bibinfo{person}{A. Auer},
  \bibinfo{person}{D.E. Bernholdt}, \bibinfo{person}{A. Bibireata},
  \bibinfo{person}{V. Choppella}, \bibinfo{person}{D. Cociorva},
  \bibinfo{person}{Xiaoyang Gao}, \bibinfo{person}{R.J. Harrison},
  \bibinfo{person}{S. Hirata}, \bibinfo{person}{S. Krishnamoorthy},
  \bibinfo{person}{S. Krishnan}, \bibinfo{person}{Chi chung Lam},
  \bibinfo{person}{Qingda Lu}, \bibinfo{person}{M. Nooijen},
  \bibinfo{person}{R.M. Pitzer}, \bibinfo{person}{J. Ramanujam},
  \bibinfo{person}{P. Sadayappan}, {and} \bibinfo{person}{A. Sibiryakov}.}
  \bibinfo{year}{2005}\natexlab{}.
\newblock \showarticletitle{Synthesis of High-Performance Parallel Programs for
  a Class of ab Initio Quantum Chemistry Models}.
\newblock \bibinfo{journal}{\emph{Proc. IEEE}} \bibinfo{volume}{93},
  \bibinfo{number}{2} (\bibinfo{year}{2005}), \bibinfo{pages}{276--292}.
\newblock
\urldef\tempurl%
\url{https://doi.org/10.1109/JPROC.2004.840311}
\showDOI{\tempurl}


\bibitem[\protect\citeauthoryear{Bondhugula}{Bondhugula}{2013}]%
        {bondugula}
\bibfield{author}{\bibinfo{person}{Uday Bondhugula}.}
  \bibinfo{year}{2013}\natexlab{}.
\newblock \showarticletitle{Compiling Affine Loop Nests for Distributed-Memory
  Parallel Architectures}. In \bibinfo{booktitle}{\emph{Proceedings of the
  International Conference on High Performance Computing, Networking, Storage
  and Analysis}} (Denver, Colorado) \emph{(\bibinfo{series}{SC '13})}.
  \bibinfo{publisher}{Association for Computing Machinery},
  \bibinfo{address}{New York, NY, USA}, Article \bibinfo{articleno}{33},
  \bibinfo{numpages}{12}~pages.
\newblock
\showISBNx{9781450323789}
\urldef\tempurl%
\url{https://doi.org/10.1145/2503210.2503289}
\showDOI{\tempurl}


\bibitem[\protect\citeauthoryear{Cannon}{Cannon}{1969}]%
        {cannon}
\bibfield{author}{\bibinfo{person}{Lynn~Elliot Cannon}.}
  \bibinfo{year}{1969}\natexlab{}.
\newblock \emph{\bibinfo{title}{A Cellular Computer to Implement the Kalman
  Filter Algorithm}}.
\newblock \bibinfo{thesistype}{Ph.D. Dissertation}. \bibinfo{address}{USA}.
\newblock
\newblock
\shownote{AAI7010025.}


\bibitem[\protect\citeauthoryear{Chamberlain, Callahan, and Zima}{Chamberlain
  et~al\mbox{.}}{2007}]%
        {chapel}
\bibfield{author}{\bibinfo{person}{B.L. Chamberlain}, \bibinfo{person}{D.
  Callahan}, {and} \bibinfo{person}{H.P. Zima}.}
  \bibinfo{year}{2007}\natexlab{}.
\newblock \showarticletitle{Parallel Programmability and the Chapel Language}.
\newblock \bibinfo{journal}{\emph{The International Journal of High Performance
  Computing Applications}} \bibinfo{volume}{21}, \bibinfo{number}{3}
  (\bibinfo{year}{2007}), \bibinfo{pages}{291--312}.
\newblock
\urldef\tempurl%
\url{https://doi.org/10.1177/1094342007078442}
\showDOI{\tempurl}
\showeprint{https://doi.org/10.1177/1094342007078442}


\bibitem[\protect\citeauthoryear{Chen, Moreau, Jiang, Zheng, Yan, Cowan, Shen,
  Wang, Hu, Ceze, Guestrin, and Krishnamurthy}{Chen et~al\mbox{.}}{2018}]%
        {TVM}
\bibfield{author}{\bibinfo{person}{Tianqi Chen}, \bibinfo{person}{Thierry
  Moreau}, \bibinfo{person}{Ziheng Jiang}, \bibinfo{person}{Lianmin Zheng},
  \bibinfo{person}{Eddie Yan}, \bibinfo{person}{Meghan Cowan},
  \bibinfo{person}{Haichen Shen}, \bibinfo{person}{Leyuan Wang},
  \bibinfo{person}{Yuwei Hu}, \bibinfo{person}{Luis Ceze},
  \bibinfo{person}{Carlos Guestrin}, {and} \bibinfo{person}{Arvind
  Krishnamurthy}.} \bibinfo{year}{2018}\natexlab{}.
\newblock \bibinfo{title}{TVM: An Automated End-to-End Optimizing Compiler for
  Deep Learning}.
\newblock
\newblock
\showeprint[arxiv]{1802.04799}~[cs.LG]


\bibitem[\protect\citeauthoryear{Choi, Dongarra, Pozo, and Walker}{Choi
  et~al\mbox{.}}{1992}]%
        {scalapack}
\bibfield{author}{\bibinfo{person}{J. Choi}, \bibinfo{person}{J.J. Dongarra},
  \bibinfo{person}{R. Pozo}, {and} \bibinfo{person}{D.W. Walker}.}
  \bibinfo{year}{1992}\natexlab{}.
\newblock \showarticletitle{ScaLAPACK: a scalable linear algebra library for
  distributed memory concurrent computers}. In
  \bibinfo{booktitle}{\emph{[Proceedings 1992] The Fourth Symposium on the
  Frontiers of Massively Parallel Computation}}. \bibinfo{pages}{120--127}.
\newblock
\urldef\tempurl%
\url{https://doi.org/10.1109/FMPC.1992.234898}
\showDOI{\tempurl}


\bibitem[\protect\citeauthoryear{Choi, Walker, and Dongarra}{Choi
  et~al\mbox{.}}{1994}]%
        {pumma}
\bibfield{author}{\bibinfo{person}{Jaeyoung Choi}, \bibinfo{person}{David~W.
  Walker}, {and} \bibinfo{person}{Jack~J. Dongarra}.}
  \bibinfo{year}{1994}\natexlab{}.
\newblock \showarticletitle{Pumma: Parallel universal matrix multiplication
  algorithms on distributed memory concurrent computers}.
\newblock \bibinfo{journal}{\emph{Concurrency: Practice and Experience}}
  \bibinfo{volume}{6}, \bibinfo{number}{7} (\bibinfo{year}{1994}),
  \bibinfo{pages}{543--570}.
\newblock
\urldef\tempurl%
\url{https://doi.org/10.1002/cpe.4330060702}
\showDOI{\tempurl}
\showeprint{https://onlinelibrary.wiley.com/doi/pdf/10.1002/cpe.4330060702}


\bibitem[\protect\citeauthoryear{Deitz, Chamberlain, and Snyder}{Deitz
  et~al\mbox{.}}{2004}]%
        {zpl}
\bibfield{author}{\bibinfo{person}{S.J. Deitz}, \bibinfo{person}{B.L.
  Chamberlain}, {and} \bibinfo{person}{L. Snyder}.}
  \bibinfo{year}{2004}\natexlab{}.
\newblock \showarticletitle{Abstractions for dynamic data distribution}. In
  \bibinfo{booktitle}{\emph{Ninth International Workshop on High-Level Parallel
  Programming Models and Supportive Environments, 2004. Proceedings.}}
  \bibinfo{pages}{42--51}.
\newblock
\urldef\tempurl%
\url{https://doi.org/10.1109/HIPS.2004.1299189}
\showDOI{\tempurl}


\bibitem[\protect\citeauthoryear{Demmel, Eliahu, Fox, Kamil, Lipshitz,
  Schwartz, and Spillinger}{Demmel et~al\mbox{.}}{2013}]%
        {carma}
\bibfield{author}{\bibinfo{person}{James Demmel}, \bibinfo{person}{David
  Eliahu}, \bibinfo{person}{Armando Fox}, \bibinfo{person}{Shoaib Kamil},
  \bibinfo{person}{Benjamin Lipshitz}, \bibinfo{person}{Oded Schwartz}, {and}
  \bibinfo{person}{Omer Spillinger}.} \bibinfo{year}{2013}\natexlab{}.
\newblock \showarticletitle{Communication-Optimal Parallel Recursive
  Rectangular Matrix Multiplication}. In \bibinfo{booktitle}{\emph{2013 IEEE
  27th International Symposium on Parallel and Distributed Processing}}.
  \bibinfo{pages}{261--272}.
\newblock
\urldef\tempurl%
\url{https://doi.org/10.1109/IPDPS.2013.80}
\showDOI{\tempurl}


\bibitem[\protect\citeauthoryear{Denniston, Kamil, and Amarasinghe}{Denniston
  et~al\mbox{.}}{2016}]%
        {dist-halide}
\bibfield{author}{\bibinfo{person}{Tyler Denniston}, \bibinfo{person}{Shoaib
  Kamil}, {and} \bibinfo{person}{Saman Amarasinghe}.}
  \bibinfo{year}{2016}\natexlab{}.
\newblock \showarticletitle{Distributed Halide}. In
  \bibinfo{booktitle}{\emph{Proceedings of the 21st ACM SIGPLAN Symposium on
  Principles and Practice of Parallel Programming}} (Barcelona, Spain)
  \emph{(\bibinfo{series}{PPoPP '16})}. \bibinfo{publisher}{Association for
  Computing Machinery}, \bibinfo{address}{New York, NY, USA}, Article
  \bibinfo{articleno}{5}, \bibinfo{numpages}{12}~pages.
\newblock
\showISBNx{9781450340922}
\urldef\tempurl%
\url{https://doi.org/10.1145/2851141.2851157}
\showDOI{\tempurl}


\bibitem[\protect\citeauthoryear{Dryden, Maruyama, Moon, Benson, Snir, and
  Van~Essen}{Dryden et~al\mbox{.}}{2019}]%
        {snir-notation}
\bibfield{author}{\bibinfo{person}{Nikoli Dryden}, \bibinfo{person}{Naoya
  Maruyama}, \bibinfo{person}{Tim Moon}, \bibinfo{person}{Tom Benson},
  \bibinfo{person}{Marc Snir}, {and} \bibinfo{person}{Brian Van~Essen}.}
  \bibinfo{year}{2019}\natexlab{}.
\newblock \showarticletitle{Channel and Filter Parallelism for Large-Scale CNN
  Training}. In \bibinfo{booktitle}{\emph{Proceedings of the International
  Conference for High Performance Computing, Networking, Storage and Analysis}}
  (Denver, Colorado) \emph{(\bibinfo{series}{SC '19})}.
  \bibinfo{publisher}{Association for Computing Machinery},
  \bibinfo{address}{New York, NY, USA}, Article \bibinfo{articleno}{10},
  \bibinfo{numpages}{20}~pages.
\newblock
\showISBNx{9781450362290}
\urldef\tempurl%
\url{https://doi.org/10.1145/3295500.3356207}
\showDOI{\tempurl}


\bibitem[\protect\citeauthoryear{Fatahalian, Horn, Knight, Leem, Houston, Park,
  Erez, Ren, Aiken, Dally, and Hanrahan}{Fatahalian et~al\mbox{.}}{2006}]%
        {Sequoia}
\bibfield{author}{\bibinfo{person}{Kayvon Fatahalian},
  \bibinfo{person}{Daniel~Reiter Horn}, \bibinfo{person}{Timothy~J. Knight},
  \bibinfo{person}{Larkhoon Leem}, \bibinfo{person}{Mike Houston},
  \bibinfo{person}{Ji~Young Park}, \bibinfo{person}{Mattan Erez},
  \bibinfo{person}{Manman Ren}, \bibinfo{person}{Alex Aiken},
  \bibinfo{person}{William~J. Dally}, {and} \bibinfo{person}{Pat Hanrahan}.}
  \bibinfo{year}{2006}\natexlab{}.
\newblock \showarticletitle{Sequoia: Programming the Memory Hierarchy}. In
  \bibinfo{booktitle}{\emph{Proceedings of the 2006 ACM/IEEE Conference on
  Supercomputing}} (Tampa, Florida) \emph{(\bibinfo{series}{SC '06})}.
  \bibinfo{publisher}{Association for Computing Machinery},
  \bibinfo{address}{New York, NY, USA}, \bibinfo{pages}{83–es}.
\newblock
\showISBNx{0769527000}
\urldef\tempurl%
\url{https://doi.org/10.1145/1188455.1188543}
\showDOI{\tempurl}


\bibitem[\protect\citeauthoryear{Ihadadene}{Ihadadene}{2019}]%
        {tiramisu-comm}
\bibfield{author}{\bibinfo{person}{Thinhinane Ihadadene}.}
  \bibinfo{year}{2019}\natexlab{}.
\newblock \bibinfo{title}{Generating Communication Code Automatically for
  Distributed Programs in Tiramisu}.
\newblock
\newblock


\bibitem[\protect\citeauthoryear{Jankov, Yuan, Luo, and Jermaine}{Jankov
  et~al\mbox{.}}{2021}]%
        {jermaine-dist-tensors}
\bibfield{author}{\bibinfo{person}{Dimitrije Jankov}, \bibinfo{person}{Binhang
  Yuan}, \bibinfo{person}{Shangyu Luo}, {and} \bibinfo{person}{Chris
  Jermaine}.} \bibinfo{year}{2021}\natexlab{}.
\newblock \showarticletitle{Distributed Numerical and Machine Learning
  Computations via Two-Phase Execution of Aggregated Join Trees}.
\newblock \bibinfo{journal}{\emph{Proc. VLDB Endow.}} \bibinfo{volume}{14},
  \bibinfo{number}{7} (\bibinfo{date}{March} \bibinfo{year}{2021}),
  \bibinfo{pages}{1228–1240}.
\newblock
\showISSN{2150-8097}
\urldef\tempurl%
\url{https://doi.org/10.14778/3450980.3450991}
\showDOI{\tempurl}


\bibitem[\protect\citeauthoryear{Kim, Sukumaran-Rajam, Thumma, Krishnamoorthy,
  Panyala, Pouchet, Rountev, and Sadayappan}{Kim et~al\mbox{.}}{2019}]%
        {cogent}
\bibfield{author}{\bibinfo{person}{Jinsung Kim}, \bibinfo{person}{Aravind
  Sukumaran-Rajam}, \bibinfo{person}{Vineeth Thumma}, \bibinfo{person}{Sriram
  Krishnamoorthy}, \bibinfo{person}{Ajay Panyala}, \bibinfo{person}{Louis-Noël
  Pouchet}, \bibinfo{person}{Atanas Rountev}, {and} \bibinfo{person}{P.
  Sadayappan}.} \bibinfo{year}{2019}\natexlab{}.
\newblock \showarticletitle{A Code Generator for High-Performance Tensor
  Contractions on GPUs}. In \bibinfo{booktitle}{\emph{2019 IEEE/ACM
  International Symposium on Code Generation and Optimization (CGO)}}.
  \bibinfo{pages}{85--95}.
\newblock
\urldef\tempurl%
\url{https://doi.org/10.1109/CGO.2019.8661182}
\showDOI{\tempurl}


\bibitem[\protect\citeauthoryear{Kjolstad, Ahrens, Kamil, and
  Amarasinghe}{Kjolstad et~al\mbox{.}}{2019}]%
        {taco_workspaces}
\bibfield{author}{\bibinfo{person}{Fredrik Kjolstad}, \bibinfo{person}{Peter
  Ahrens}, \bibinfo{person}{Shoaib Kamil}, {and} \bibinfo{person}{Saman
  Amarasinghe}.} \bibinfo{year}{2019}\natexlab{}.
\newblock \showarticletitle{Tensor Algebra Compilation with Workspaces}. In
  \bibinfo{booktitle}{\emph{2019 IEEE/ACM International Symposium on Code
  Generation and Optimization (CGO)}}. \bibinfo{pages}{180--192}.
\newblock
\urldef\tempurl%
\url{https://doi.org/10.1109/CGO.2019.8661185}
\showDOI{\tempurl}


\bibitem[\protect\citeauthoryear{Kjolstad, Kamil, Chou, Lugato, and
  Amarasinghe}{Kjolstad et~al\mbox{.}}{2017}]%
        {taco}
\bibfield{author}{\bibinfo{person}{Fredrik Kjolstad}, \bibinfo{person}{Shoaib
  Kamil}, \bibinfo{person}{Stephen Chou}, \bibinfo{person}{David Lugato}, {and}
  \bibinfo{person}{Saman Amarasinghe}.} \bibinfo{year}{2017}\natexlab{}.
\newblock \showarticletitle{The Tensor Algebra Compiler}.
\newblock \bibinfo{journal}{\emph{Proc. ACM Program. Lang.}}
  \bibinfo{volume}{1}, \bibinfo{number}{OOPSLA}, Article
  \bibinfo{articleno}{77} (\bibinfo{date}{Oct.} \bibinfo{year}{2017}),
  \bibinfo{numpages}{29}~pages.
\newblock
\urldef\tempurl%
\url{https://doi.org/10.1145/3133901}
\showDOI{\tempurl}


\bibitem[\protect\citeauthoryear{Kolda and Bader}{Kolda and Bader}{2009}]%
        {baderkolda}
\bibfield{author}{\bibinfo{person}{Tamara~G. Kolda} {and}
  \bibinfo{person}{Brett~W. Bader}.} \bibinfo{year}{2009}\natexlab{}.
\newblock \showarticletitle{Tensor Decompositions and Applications}.
\newblock \bibinfo{journal}{\emph{SIAM Rev.}} \bibinfo{volume}{51},
  \bibinfo{number}{3} (\bibinfo{date}{Aug.} \bibinfo{year}{2009}),
  \bibinfo{pages}{455–500}.
\newblock
\showISSN{0036-1445}
\urldef\tempurl%
\url{https://doi.org/10.1137/07070111X}
\showDOI{\tempurl}


\bibitem[\protect\citeauthoryear{Kwasniewski}{Kwasniewski}{[n.d.]}]%
        {gregory_comm}
\bibfield{author}{\bibinfo{person}{Grzegorz Kwasniewski}.}
  \bibinfo{year}{[n.d.]}\natexlab{}.
\newblock \bibinfo{howpublished}{personal communication}.
\newblock


\bibitem[\protect\citeauthoryear{Kwasniewski, Kabi\'{c}, Besta, VandeVondele,
  Solc\`{a}, and Hoefler}{Kwasniewski et~al\mbox{.}}{2019}]%
        {cosma}
\bibfield{author}{\bibinfo{person}{Grzegorz Kwasniewski},
  \bibinfo{person}{Marko Kabi\'{c}}, \bibinfo{person}{Maciej Besta},
  \bibinfo{person}{Joost VandeVondele}, \bibinfo{person}{Raffaele Solc\`{a}},
  {and} \bibinfo{person}{Torsten Hoefler}.} \bibinfo{year}{2019}\natexlab{}.
\newblock \showarticletitle{Red-Blue Pebbling Revisited: Near Optimal Parallel
  Matrix-Matrix Multiplication}. In \bibinfo{booktitle}{\emph{Proceedings of
  the International Conference for High Performance Computing, Networking,
  Storage and Analysis}} (Denver, Colorado) \emph{(\bibinfo{series}{SC '19})}.
  \bibinfo{publisher}{Association for Computing Machinery},
  \bibinfo{address}{New York, NY, USA}, Article \bibinfo{articleno}{24},
  \bibinfo{numpages}{22}~pages.
\newblock
\showISBNx{9781450362290}
\urldef\tempurl%
\url{https://doi.org/10.1145/3295500.3356181}
\showDOI{\tempurl}


\bibitem[\protect\citeauthoryear{LLNL}{LLNL}{2021}]%
        {lassen}
\bibfield{author}{\bibinfo{person}{LLNL}.} \bibinfo{year}{2021}\natexlab{}.
\newblock \bibinfo{title}{Lassen}.
\newblock
\newblock
\urldef\tempurl%
\url{https://hpc.llnl.gov/hardware/platforms/lassen}
\showURL{%
\tempurl}


\bibitem[\protect\citeauthoryear{Loveman}{Loveman}{1993}]%
        {high-performance-fortran}
\bibfield{author}{\bibinfo{person}{D.B. Loveman}.}
  \bibinfo{year}{1993}\natexlab{}.
\newblock \showarticletitle{High performance Fortran}.
\newblock \bibinfo{journal}{\emph{IEEE Parallel Distributed Technology: Systems
  Applications}} \bibinfo{volume}{1}, \bibinfo{number}{1}
  (\bibinfo{year}{1993}), \bibinfo{pages}{25--42}.
\newblock
\urldef\tempurl%
\url{https://doi.org/10.1109/88.219857}
\showDOI{\tempurl}


\bibitem[\protect\citeauthoryear{Luo, Gao, Gubanov, Perez, and Jermaine}{Luo
  et~al\mbox{.}}{2017}]%
        {jermaine-dist-lin-alg}
\bibfield{author}{\bibinfo{person}{Shangyu Luo}, \bibinfo{person}{Zekai~J.
  Gao}, \bibinfo{person}{Michael Gubanov}, \bibinfo{person}{Luis~L. Perez},
  {and} \bibinfo{person}{Christopher Jermaine}.}
  \bibinfo{year}{2017}\natexlab{}.
\newblock \showarticletitle{Scalable Linear Algebra on a Relational Database
  System}. In \bibinfo{booktitle}{\emph{2017 IEEE 33rd International Conference
  on Data Engineering (ICDE)}}. \bibinfo{pages}{523--534}.
\newblock
\urldef\tempurl%
\url{https://doi.org/10.1109/ICDE.2017.108}
\showDOI{\tempurl}


\bibitem[\protect\citeauthoryear{Ragan-Kelley, Barnes, Adams, Paris, Durand,
  and Amarasinghe}{Ragan-Kelley et~al\mbox{.}}{2013}]%
        {halide}
\bibfield{author}{\bibinfo{person}{Jonathan Ragan-Kelley},
  \bibinfo{person}{Connelly Barnes}, \bibinfo{person}{Andrew Adams},
  \bibinfo{person}{Sylvain Paris}, \bibinfo{person}{Fr\'{e}do Durand}, {and}
  \bibinfo{person}{Saman Amarasinghe}.} \bibinfo{year}{2013}\natexlab{}.
\newblock \showarticletitle{Halide: A Language and Compiler for Optimizing
  Parallelism, Locality, and Recomputation in Image Processing Pipelines}.
\newblock \bibinfo{journal}{\emph{SIGPLAN Not.}} \bibinfo{volume}{48},
  \bibinfo{number}{6} (\bibinfo{date}{June} \bibinfo{year}{2013}),
  \bibinfo{pages}{519–530}.
\newblock
\showISSN{0362-1340}
\urldef\tempurl%
\url{https://doi.org/10.1145/2499370.2462176}
\showDOI{\tempurl}


\bibitem[\protect\citeauthoryear{Schatz}{Schatz}{2015}]%
        {schatzthesis}
\bibfield{author}{\bibinfo{person}{Martin~Daniel Schatz}.}
  \bibinfo{year}{2015}\natexlab{}.
\newblock \emph{\bibinfo{title}{Distributed Tensor Computations: Formalizing
  Distributions, Redistributions, and Algorithm Derivation}}.
\newblock \bibinfo{thesistype}{Ph.D. Dissertation}. \bibinfo{address}{USA}.
\newblock


\bibitem[\protect\citeauthoryear{Schatz, Geijn, and Poulson}{Schatz
  et~al\mbox{.}}{2016}]%
        {elementalMM}
\bibfield{author}{\bibinfo{person}{Martin~D. Schatz},
  \bibinfo{person}{Robert~A. Geijn}, {and} \bibinfo{person}{Jack Poulson}.}
  \bibinfo{year}{2016}\natexlab{}.
\newblock \showarticletitle{Parallel Matrix Multiplication: A Systematic
  Journey}.
\newblock \bibinfo{journal}{\emph{SIAM J. Sci. Comput.}}  \bibinfo{volume}{38}
  (\bibinfo{year}{2016}).
\newblock


\bibitem[\protect\citeauthoryear{Senanayake, Hong, Wang, Wilson, Chou, Kamil,
  Amarasinghe, and Kjolstad}{Senanayake et~al\mbox{.}}{2020}]%
        {taco_scheduling}
\bibfield{author}{\bibinfo{person}{Ryan Senanayake}, \bibinfo{person}{Changwan
  Hong}, \bibinfo{person}{Ziheng Wang}, \bibinfo{person}{Amalee Wilson},
  \bibinfo{person}{Stephen Chou}, \bibinfo{person}{Shoaib Kamil},
  \bibinfo{person}{Saman Amarasinghe}, {and} \bibinfo{person}{Fredrik
  Kjolstad}.} \bibinfo{year}{2020}\natexlab{}.
\newblock \showarticletitle{A Sparse Iteration Space Transformation Framework
  for Sparse Tensor Algebra}.
\newblock \bibinfo{journal}{\emph{Proc. ACM Program. Lang.}}
  \bibinfo{volume}{4}, \bibinfo{number}{OOPSLA}, Article
  \bibinfo{articleno}{158} (\bibinfo{date}{Nov.} \bibinfo{year}{2020}),
  \bibinfo{numpages}{30}~pages.
\newblock
\urldef\tempurl%
\url{https://doi.org/10.1145/3428226}
\showDOI{\tempurl}


\bibitem[\protect\citeauthoryear{Solomonik and Demmel}{Solomonik and
  Demmel}{2011}]%
        {solomonikMM}
\bibfield{author}{\bibinfo{person}{Edgar Solomonik} {and}
  \bibinfo{person}{James Demmel}.} \bibinfo{year}{2011}\natexlab{}.
\newblock \showarticletitle{Communication-Optimal Parallel 2.5D Matrix
  Multiplication and LU Factorization Algorithms}. In
  \bibinfo{booktitle}{\emph{Euro-Par 2011 Parallel Processing}},
  \bibfield{editor}{\bibinfo{person}{Emmanuel Jeannot},
  \bibinfo{person}{Raymond Namyst}, {and} \bibinfo{person}{Jean Roman}} (Eds.).
  \bibinfo{publisher}{Springer Berlin Heidelberg}, \bibinfo{address}{Berlin,
  Heidelberg}, \bibinfo{pages}{90--109}.
\newblock
\showISBNx{978-3-642-23397-5}


\bibitem[\protect\citeauthoryear{Solomonik, Matthews, Hammond, Stanton, and
  Demmel}{Solomonik et~al\mbox{.}}{2014}]%
        {ctf}
\bibfield{author}{\bibinfo{person}{Edgar Solomonik}, \bibinfo{person}{Devin
  Matthews}, \bibinfo{person}{Jeff~R. Hammond}, \bibinfo{person}{John~F.
  Stanton}, {and} \bibinfo{person}{James Demmel}.}
  \bibinfo{year}{2014}\natexlab{}.
\newblock \showarticletitle{A massively parallel tensor contraction framework
  for coupled-cluster computations}.
\newblock \bibinfo{journal}{\emph{J. Parallel and Distrib. Comput.}}
  \bibinfo{volume}{74}, \bibinfo{number}{12} (\bibinfo{year}{2014}),
  \bibinfo{pages}{3176--3190}.
\newblock
\showISSN{0743-7315}
\urldef\tempurl%
\url{https://doi.org/10.1016/j.jpdc.2014.06.002}
\showDOI{\tempurl}
\newblock
\shownote{Domain-Specific Languages and High-Level Frameworks for
  High-Performance Computing.}


\bibitem[\protect\citeauthoryear{van~de Geijn and Watts}{van~de Geijn and
  Watts}{1995}]%
        {summa}
\bibfield{author}{\bibinfo{person}{Robert~A. van~de Geijn} {and}
  \bibinfo{person}{Jerrell Watts}.} \bibinfo{year}{1995}\natexlab{}.
\newblock \bibinfo{booktitle}{\emph{SUMMA: Scalable Universal Matrix
  Multiplication Algorithm}}.
\newblock \bibinfo{type}{{T}echnical {R}eport}. \bibinfo{address}{USA}.
\newblock


\bibitem[\protect\citeauthoryear{Vasilache, Zinenko, Theodoridis, Goyal,
  DeVito, Moses, Verdoolaege, Adams, and Cohen}{Vasilache
  et~al\mbox{.}}{2018}]%
        {tensorcomprehensions}
\bibfield{author}{\bibinfo{person}{Nicolas Vasilache},
  \bibinfo{person}{Oleksandr Zinenko}, \bibinfo{person}{Theodoros Theodoridis},
  \bibinfo{person}{Priya Goyal}, \bibinfo{person}{Zachary DeVito},
  \bibinfo{person}{William~S. Moses}, \bibinfo{person}{Sven Verdoolaege},
  \bibinfo{person}{Andrew Adams}, {and} \bibinfo{person}{Albert Cohen}.}
  \bibinfo{year}{2018}\natexlab{}.
\newblock \bibinfo{title}{Tensor Comprehensions: Framework-Agnostic
  High-Performance Machine Learning Abstractions}.
\newblock
\newblock
\showeprint[arxiv]{1802.04730}~[cs.PL]


\bibitem[\protect\citeauthoryear{Zhang, Yang, Baghdadi, Kamil, Shun, and
  Amarasinghe}{Zhang et~al\mbox{.}}{2018}]%
        {graphit}
\bibfield{author}{\bibinfo{person}{Yunming Zhang}, \bibinfo{person}{Mengjiao
  Yang}, \bibinfo{person}{Riyadh Baghdadi}, \bibinfo{person}{Shoaib Kamil},
  \bibinfo{person}{Julian Shun}, {and} \bibinfo{person}{Saman Amarasinghe}.}
  \bibinfo{year}{2018}\natexlab{}.
\newblock \showarticletitle{GraphIt: A High-Performance Graph DSL}.
\newblock \bibinfo{journal}{\emph{Proc. ACM Program. Lang.}}
  \bibinfo{volume}{2}, \bibinfo{number}{OOPSLA}, Article
  \bibinfo{articleno}{121} (\bibinfo{date}{Oct.} \bibinfo{year}{2018}),
  \bibinfo{numpages}{30}~pages.
\newblock
\urldef\tempurl%
\url{https://doi.org/10.1145/3276491}
\showDOI{\tempurl}


\end{thebibliography}

\end{document}